\documentclass[a4paper,fleqn,usenatbib]{mnras}

\usepackage{newtxtext,newtxmath}
\usepackage[T1]{fontenc}
\usepackage{ae,aecompl}
\usepackage{graphicx}	
\usepackage{amsmath}	
\usepackage{amssymb}	
\usepackage{color}

\newcommand{\pd}{\mathrm{\partial}}

\title[]{Three-dimensional core-collapse supernova simulations of massive and rotating progenitors}

\author[Jade Powell \& Bernhard  M\"uller]{
  Jade Powell$^{1}$\thanks{E-mail: dr.jade.powell@gmail.com}
  and
Bernhard M\"uller$^{2}$
\\
$^{1}$OzGrav, Centre for Astrophysics and Supercomputing, Swinburne University of Technology, Hawthorn, VIC 3122, Australia.\\
$^{2}$Monash Centre for Astrophysics, School of Physics and Astronomy, Monash University, VIC 3800, Australia.\\
}

\date{}

\pubyear{2020}

\begin{document}
\label{firstpage}
\pagerange{\pageref{firstpage}--\pageref{lastpage}}
\maketitle

\begin{abstract}
We present three-dimensional simulations of the core-collapse of
massive rotating and non-rotating progenitors performed with the
general relativistic neutrino hydrodynamics code \textsc{CoCoNuT-FMT}
and analyse their explosion properties and gravitational-wave
signals. The progenitor models include Wolf-Rayet stars with initial
helium star masses of $39\,M_{\odot}$ and $20\,M_{\odot}$, and an
$18\,M_{\odot}$ red supergiant. The $39\,M_{\odot}$ model is a rapid
rotator, whereas the two other progenitors are non-rotating. Both
Wolf-Rayet models produce healthy neutrino-driven explosions, whereas
the red supergiant model fails to explode. By the end of the
simulations, the explosion energies have already reached $1.1\times
10^{51}\,\mathrm{erg}$ and $0.6\times 10^{51}\,\mathrm{erg}$ for the
$39\,M_{\odot}$ and $20\,M_{\odot}$ model, respectively. The
explosions produce neutron stars of relatively high mass, but with
modest kicks. Due to the alignment of the bipolar explosion geometry
with the rotation axis, there is a relatively small misalignment of
$30^\circ$ between the spin and the kick in the $39\,M_{\odot}$
model. In terms of gravitational-wave signals, the massive and rapidly
rotating $39\,M_{\odot}$ progenitor stands out by large
gravitational-wave amplitudes that would make it detectable out to
almost 2\,Mpc by the Einstein Telescope. For this model, we find that
rotation significantly changes the dependence of the characteristic
gravitational-wave frequency of the f-mode on the proto-neutron star
parameters compared to the non-rotating case. The other two progenitors
have considerably smaller detection distances, despite significant
low-frequency emission in the most sensitive frequency band of current
gravitational-wave detectors due to the standing accretion shock
instability in the $18\,M_{\odot}$ model.
\end{abstract}


\section{Introduction}
\label{sec:intro}

Multi-dimensional simulations of core-collapse supernovae (CCSNe) have advanced 
rapidly in recent years (See \citealt{2016PASA...33...48M,janka_16} for recent reviews), with 
many simulations focusing on three-dimensional (3D), non-rotating, neutrino-driven
explosions. The neutrino-driven explosion mechanism involves some re-absorption 
of the emitted neutrinos behind the stalled shock, which revive the shock wave 
and power the explosion \citep{2012ARNPS..62..407J, 2013RvMP...85..245B}.
Successful neutrino-driven shock revival has now been observed in multiple
3D simulations employing different codes and a range
of progenitor parameters 
\citep[e.g.][]{melson_15b,lentz_15,takiwaki_14,2018ApJ...855L...3O,2017MNRAS.472..491M,2019MNRAS.484.3307M,
2019arXiv190904152B, 2019MNRAS.487.1178P,burrows_19b,walk_19}. 

The frequency range of the gravitational waves emitted in neutrino-driven explosions 
makes CCSNe a promising source for current ground based gravitational-wave detectors, 
such as Advanced LIGO (aLIGO; \citealp{aLIGO}) and Advanced Virgo (AdVirgo; \citealp{AdVirgo}), 
and planned future detectors like the Einstein Telescope (ET; \citealp{0264-9381-27-19-194002}).
Recent 3D simulations of neutrino-driven CCSNe have shown that the 
dominant feature of the gravitational-wave emission is due to the quadrupolar surface 
g-mode of the proto-neutron star (PNS), which is excited by the hydrodynamical 
instabilities inside and outside the PNS \citep{2009ApJ...707.1173M, 2013ApJ...766...43M, 
2013ApJ...779L..18C, 2017MNRAS.468.2032A, 2016ApJ...829L..14K}. At lower frequencies 
(below $\sim200$\,Hz), there may be further features associated with the standing accretion shock 
instability (SASI) \citep{0004-637X-584-2-971, 2006ApJ...642..401B, 2007ApJ...654.1006F}.

The majority of successful 3D neutrino-driven CCSN simulations are 
explosions of non-rotating progenitor models with zero-age main sequence (ZAMS) 
masses in the range of $10\,M_{\odot}$ to $30\,M_{\odot}$. 
To fully cover the CCSN parameter space, further 
3D simulations of higher mass stars are required. 
Whether and when such massive stars directly form 
black holes, or explode and leave massive neutron stars, or a black hole due to 
fallback, is an open question. Simulating CCSNe with larger progenitor masses  
will also enable us to further study the relationship between compactness 
and explodability found in previous studies \citep{2011ApJ...730...70O}.
Intriguingly and in contrast to \citet{2011ApJ...730...70O}, a number
of  3D neutrino-driven simulations 
\citep{chan_18,2018ApJ...855L...3O,kuroda_18,2019arXiv190904152B,burrows_19b,walk_19}
observed shock expansion in some progenitors with rather massive and compact
cores.  
From the point of view of gravitational-wave astronomy, high-mass progenitors
are particularly interesting because they are expected to explode more energetically
\citep{2016MNRAS.460..742M}, which will result in stronger gravitational-wave emission
\citep{mueller_17,2019MNRAS.487.1178P,radice_18}.
Progenitor models with large 
compactness parameters are also expected to more closely reproduce the observed 
levels of $^{56}\mathrm{Ni}$ \citep{2019MNRAS.483.3607S}.  

Further to this, as most studies of neutrino-driven explosions have used non-rotating models, the
effects of rotation on the explosion dynamics, remnant properties, and multi-messenger 
observables are not well understood. 
Most stars are thought to be slowly rather than rapidly rotating 
\citep{2014A&A...564A..27D, 2012Natur.481...55B}. However, the rotation rate of a progenitor 
star at core-collapse may have a significant impact on the characteristics and mechanism of 
the explosion, the multi-messenger observable emission, and the nature of the compact remnant. 
The first 3D simulations of neutrino-driven explosions in rotating progenitors suggest
that rotation can play a supportive role for shock revival even without
invoking magnetic fields
\citep{takiwaki_16,janka_16,summa}. In particular, a violent corotation instability
can develop for sufficiently high rotation rates \citep{takiwaki_16}.
In the presence of strong magnetic fields or fast field amplifications,
powerful magnetorotational explosions might develop, and the subsequent
evolution may lead to the launching of relativistic jets either
from a millisecond magnetar \citep{uzov_92} or from an accretion disk
in the collapsar scenario \citep{1999ApJ...524..262M}. While first 3D simulations
of such magnetorotational explosions have become possible 
\citep{winteler_12,moesta}, many uncertainties remain about the initial field strength
and topology, the initial rotation profiles, field amplification
and non-ideal effects \citep{guilet_15}, and the stability of MHD-driven jets
\citep{moesta}.

The gravitational-wave signal from the bounce and early post-bounce phase in
rapidly rotating progenitors has already been studied extensively in 2D and 3D and is
well understood \citep{ott,2007A&A...474..169C, 
2008PhRvD..78f4056D, 2010A&A...514A..51S,
2014PhRvD..90d4001A, 2017PhRvD..95f3019R,2018MNRAS.475L..91T,2019ApJ...878...13P}.
3D studies of the gravitational-wave signal from the later
post-bounce phase have recently become available \citep{2014PhRvD..89d4011K,
2018MNRAS.475L..91T, 2019arXiv190909730S, 2018arXiv181007638A}. However, some of these studies 
do not include general relativity, or are stopped 
before 300\,ms when the amplitudes of the gravitational-wave signals are still growing. These
studies have shown that rotation is expected to significantly increase the amplitude of the 
gravitational-wave emission.

In this paper, we
seek to produce long-duration, 3D simulations of very massive 
progenitors including rotation, general relativity, and multi-group
neutrino transport to better understand the progenitor regime from
which we expect the strongest gravitational-wave signals, thus
complementing our previous study in \citet{2019MNRAS.487.1178P}. 
We perform two CCSN simulations of Wolf-Rayet stars in 3D
with the neutrino hydrodynamics code \textsc{CoCoNuT-FMT}. The first is a 
rapidly rotating star with an initial helium star mass of $39\,M_{\odot}$. We show 
that this model produces a powerful neutrino-driven explosion due to rotational support without 
the aid of magnetic fields. Our second model is a Wolf-Rayet star with an initial helium 
star mass of $20\,M_{\odot}$. 

The high mass and (in one case) rapid rotation of the models allow us to 
estimate the maximum distance at which ground-based gravitational-wave detectors will be 
sensitive to neutrino-driven CCSNe. Further to this, we include a smaller $18\,M_{\odot}$  
ZAMS star that differs from our previous $18\,M_{\odot}$ model in 
\citet{2019MNRAS.487.1178P} by not including perturbations from the convective
oxygen shell. As a result, this model develops
strong SASI after collapse, which produces gravitational waves in the optimum sensitivity range 
for aLIGO and AdVirgo. We use this model to determine if the SASI increases 
detectability of gravitational-wave signals in comparison to our other convection-dominated models.

The outline of our paper is as follows: In Section \ref{sec:sim}, we describe the setup
of our simulation and outline the details of our neutrino hydrodynamics code. In Section
\ref{sec:models}, we describe in more detail our progenitor models. In Section 
\ref{sec:dynamics}, we describe the explosion dynamics including the effects of rotation,
and we discuss the properties of the PNS in Section \ref{sec:properties}. 
We show the gravitational-wave emission in Section \ref{sec:gw},
and estimate the detectability of our models. A conclusion and discussion are given in 
Section \ref{sec:conclusion}. 

\section{Simulation Methodology and Setup}
\label{sec:sim}

We carry out 3D simulations using the neutrino hydrodynamics code
\textsc{CoCoNuT-FMT}. For the equations of hydrodynamics, we use a general relativistic
finite-volume based solver in spherical polar coordinates 
\citep{2010ApJS..189..104M, 2002A&A...393..523D}. It employs the xCFC approximation 
for the space-time metric \citep{PhysRevD.79.024017}.
We use the fast multigroup neutrino transport method described in
\citet{mueller_15a}, including the updates listed in \citet{2019MNRAS.487.1178P}. 

We use a spatial resolution of 550x128x256 and 21 energy groups. We use the Lattimer 
and Swesty equation of state, with a bulk incompressibility of K=220\,MeV, 
at high densities \citep{Lattimer:1991nc}. At low density, we use an equation of state 
accounting for photons, electrons, positrons, and an ideal gas of nuclei together with
a flashing treatment for nuclear reactions \citep{rampp_02}. 

\section{Progenitor Models}
\label{sec:models}

\begin{table*}
\begin{center}
\begin{tabular}{|c|c|c|c|c|c|c|c|} 
 \hline
 Model & $\mathrm{M}_{\mathrm{He}} \,(M_{\odot})$ & $\omega_\mathrm{c}\, (\mathrm{km\, s}^{-1})$ & $\mathrm{M}_{\mathrm{by}} \,(M_{\odot})$ & $\mathrm{M}_{\mathrm{grav}} \,(M_{\odot})$ & $\mathrm{E}_{\mathrm{expl}} \,(10^{50} \mathrm{erg}) $ & $v_\mathrm{pns} \,(\mathrm{km\, s}^{-1})$ & $\alpha$ \\ 
 \hline
 m39 & 39 & 0.542 & 2.04 & 1.77 & 11.0 & 127.4 & 30.5 \\ 
 \hline
 y20 & 20 & 0 & 1.75 & 1.55 & 5.9 & 52.2 & 70.9 \\
 \hline
 s18np & 5.3 & 0 & 1.86 & 1.64 & 0 & 0 & 0 \\
 \hline
 \end{tabular}
\caption{
$\mathrm{M}_{\mathrm{He}}$ is the initial helium star mass. 
$\omega_\mathrm{c}$ is the central angular velocity at the pre-collapse stage.
${M}_{\mathrm{by}}$ in the final baryonic mass and $\mathrm{M}_{\mathrm{grav}}$ is the final gravitational mass. $\mathrm{E}_{\mathrm{expl}}$ is the final explosion energy. $v_\mathrm{pns}$ is the final kick velocity, and $\alpha$ is the angle between the spin and kick. 
}
\label{tab:properties}
\end{center}
\end{table*}

\subsection{Model m39 -- Rapidly Rotating Wolf-Rayet Star}
Our first model, m39, is a rapidly rotating
Wolf-Rayet star.
It has an initial helium star mass of $39\,M_{\odot}$, 
with $22\,M_{\odot}$ left at the time of collapse \citep{2018ApJ...858..115A}.
The model was evolved with the Modules for Experiments in Stellar 
Astrophysics (MESA) stellar evolution code \citep{2011ApJS..192....3P}. 
It rotates rapidly with an initial surface rotation velocity of 
$600\,\mathrm{kms}^{-1}$ and has a metallicity of $1/50\,Z_{\odot}$.
During the collapse phase, we evolve this model in 2D until bounce, and then map to 3D, adding random perturbations in density to break axial
symmetry.
We end the simulation of this model at 0.98\,s after core bounce.

\subsection{Model y20 -- Non-Rotating Wolf-Rayet Star}
\citet{2017MNRAS.470.3970Y} produced stellar evolution models for Wolf-Rayet 
stars with an initial helium mass range of $4-25\,M_{\odot}$ using the binary
evolution code BEC \citep{2001A&A...369..939W}. 
For our second model, y20, we use the 
$20\,M_{\odot}$, non-rotating, solar metallicity helium star from 
this study. It was evolved with the mass-loss prescription of \citet{2000A&A...360..227N},
which results in a final helium core mass of $10\,M_{\odot}$ after wind loss.
We end the simulation of this model at 1.2\,s after core bounce. 

\subsection{Model s18np -- Massive Red Supergiant Progenitor}
Our third model, s18np, is a non-rotating, solar metallicity star with a ZAMS mass 
of $18\,M_{\odot}$ \citep{2016ApJ...833..124M}. It was evolved up to the 
onset of collapse with the stellar evolution code KEPLER \citep{2017ascl.soft02007W}.
This model has been simulated previously in \citet{2017MNRAS.472..491M}, however
the gravitational-wave emission was not calculated as the PNS convection zone  
was simulated in spherical symmetry. Different from the $18\,M_{\odot}$ model that 
we simulated in \citet{2019MNRAS.487.1178P}, this model does not include 
strong pre-shock density perturbations from convective
oxygen burning. As a result, the shock 
is not revived, and strong SASI activity develops and is maintained for hundreds of milliseconds. Since the SASI produces gravitational waves in the optimum sensitivity range for 
ground based gravitational-wave detectors, this model allows us to 
explore the effects of the SASI on the detectability of the gravitational-wave signal. 
We end the simulation at 0.56\,s after core bounce after the amplitude of the 
gravitational-wave emission has decreased significantly.  

\section{Explosion Dynamics}
\label{sec:dynamics}

\begin{figure*}
\centering
\includegraphics[width=0.32\textwidth]{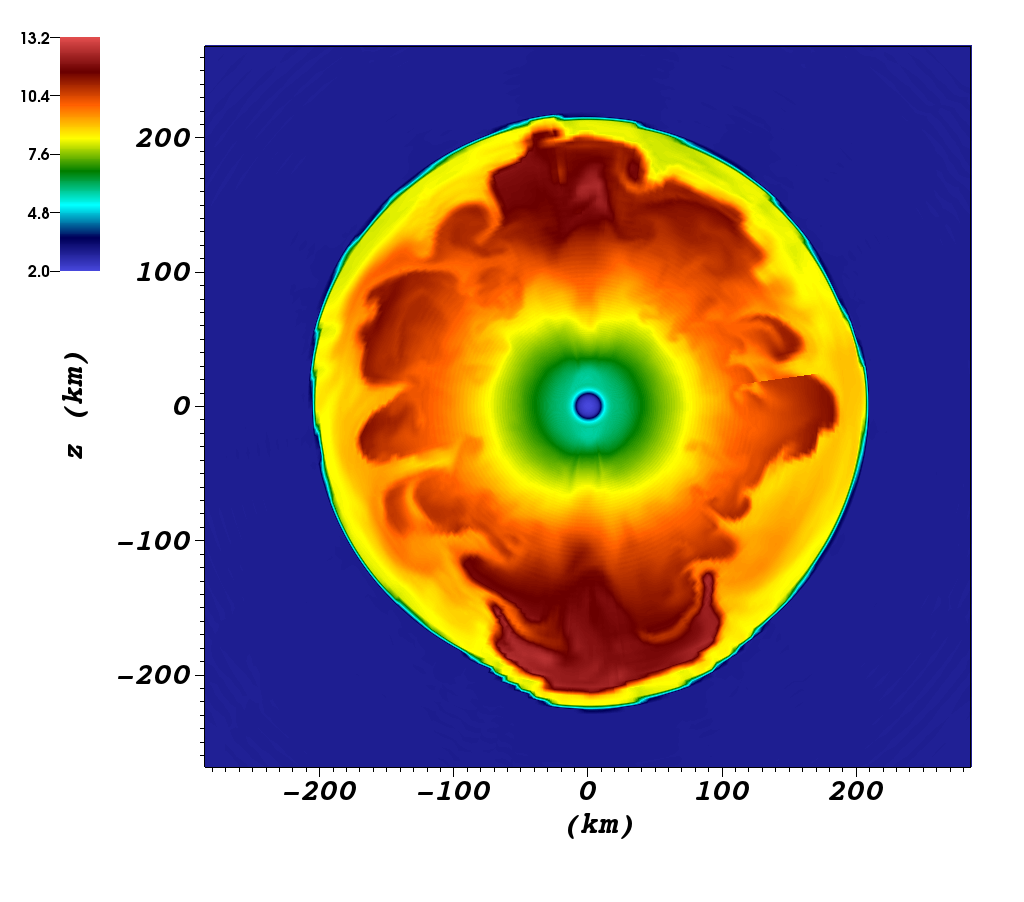}
\includegraphics[width=0.32\textwidth]{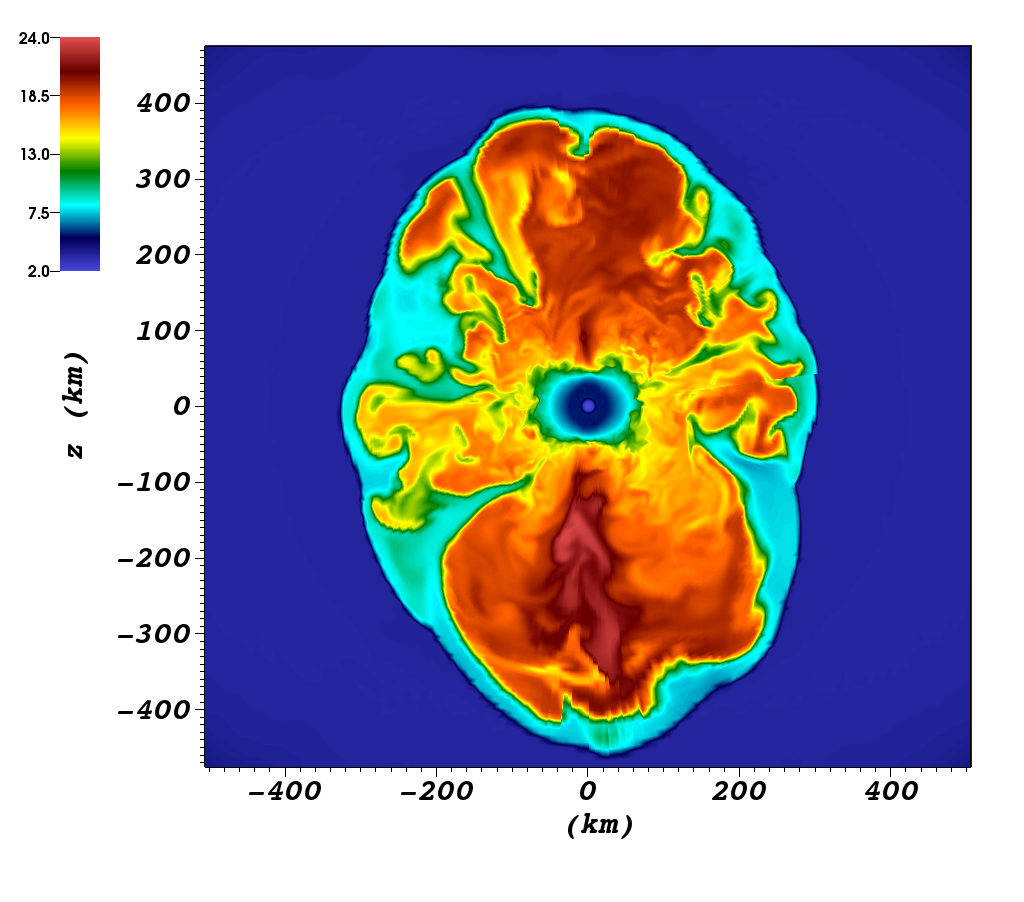}
\includegraphics[width=0.32\textwidth]{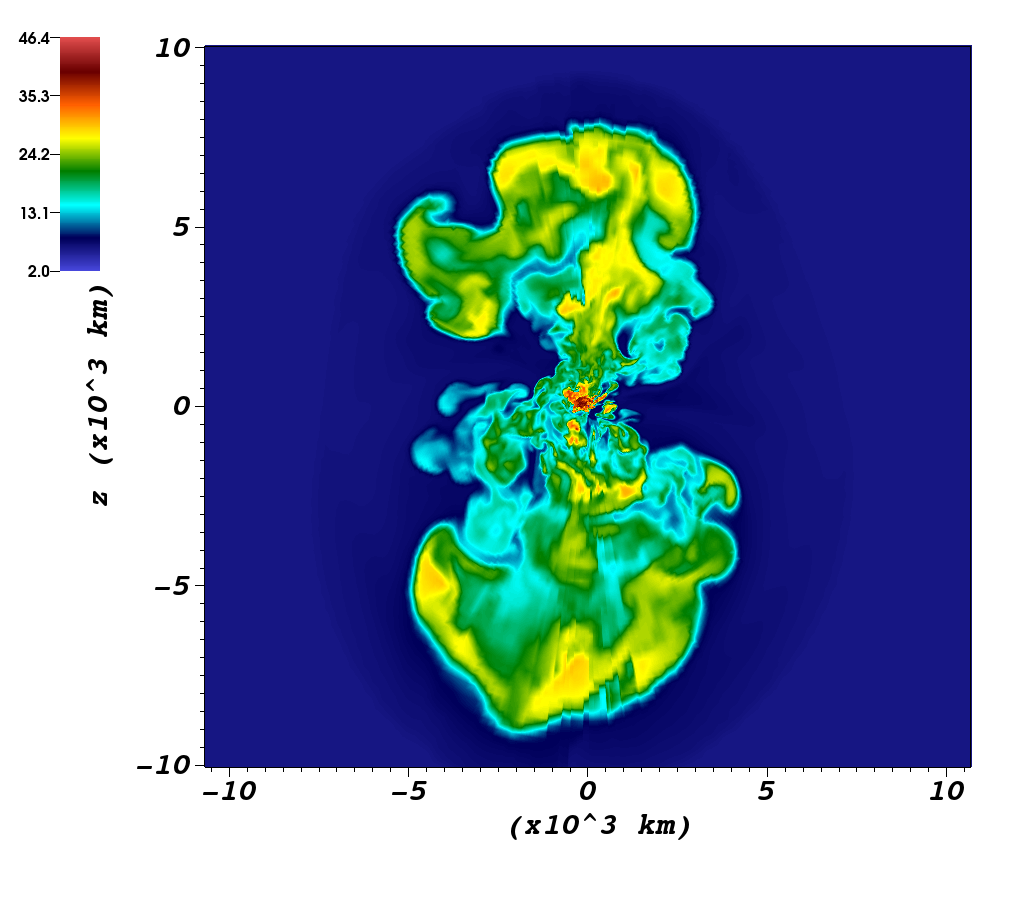}

\includegraphics[width=0.32\textwidth]{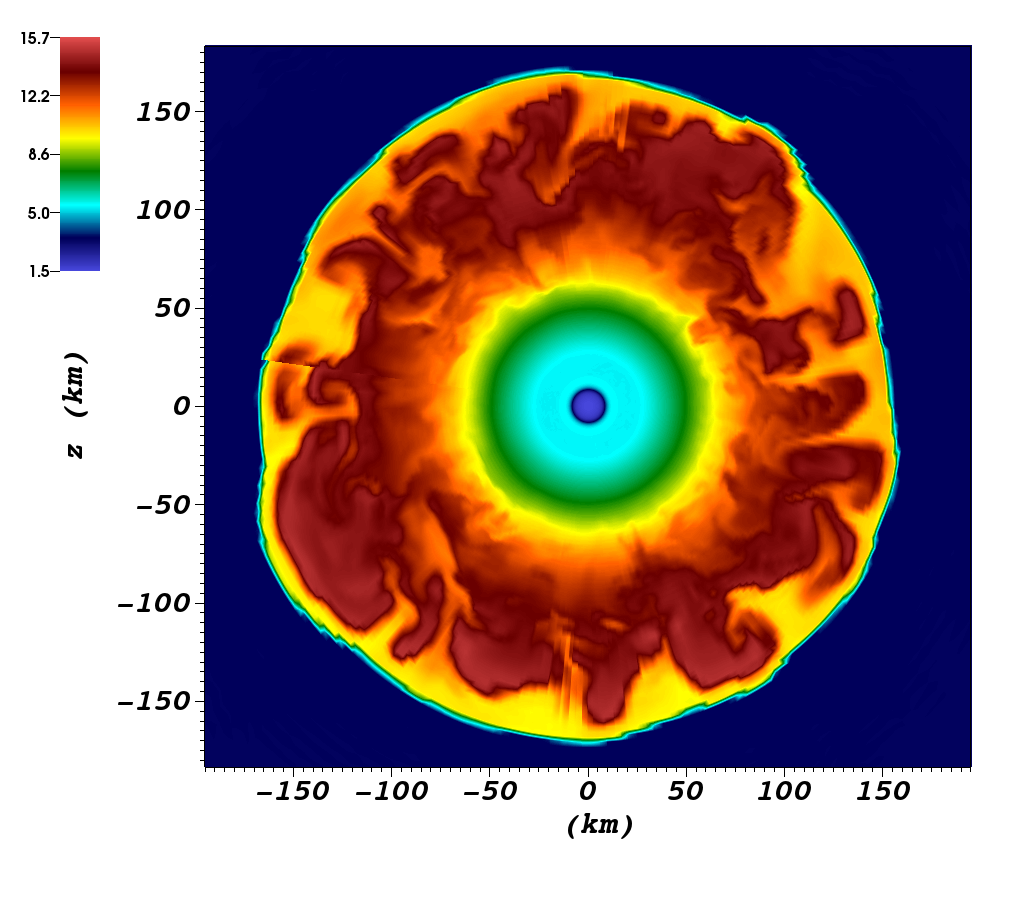}
\includegraphics[width=0.32\textwidth]{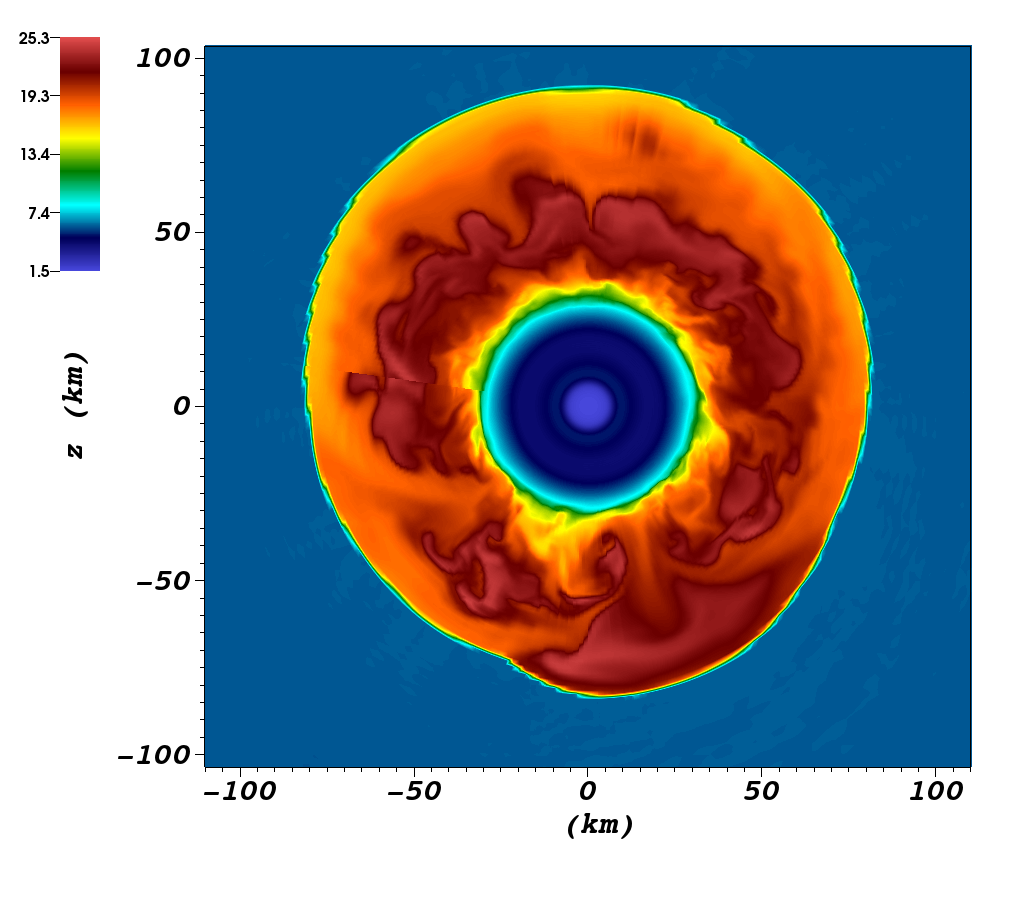}
\includegraphics[width=0.32\textwidth]{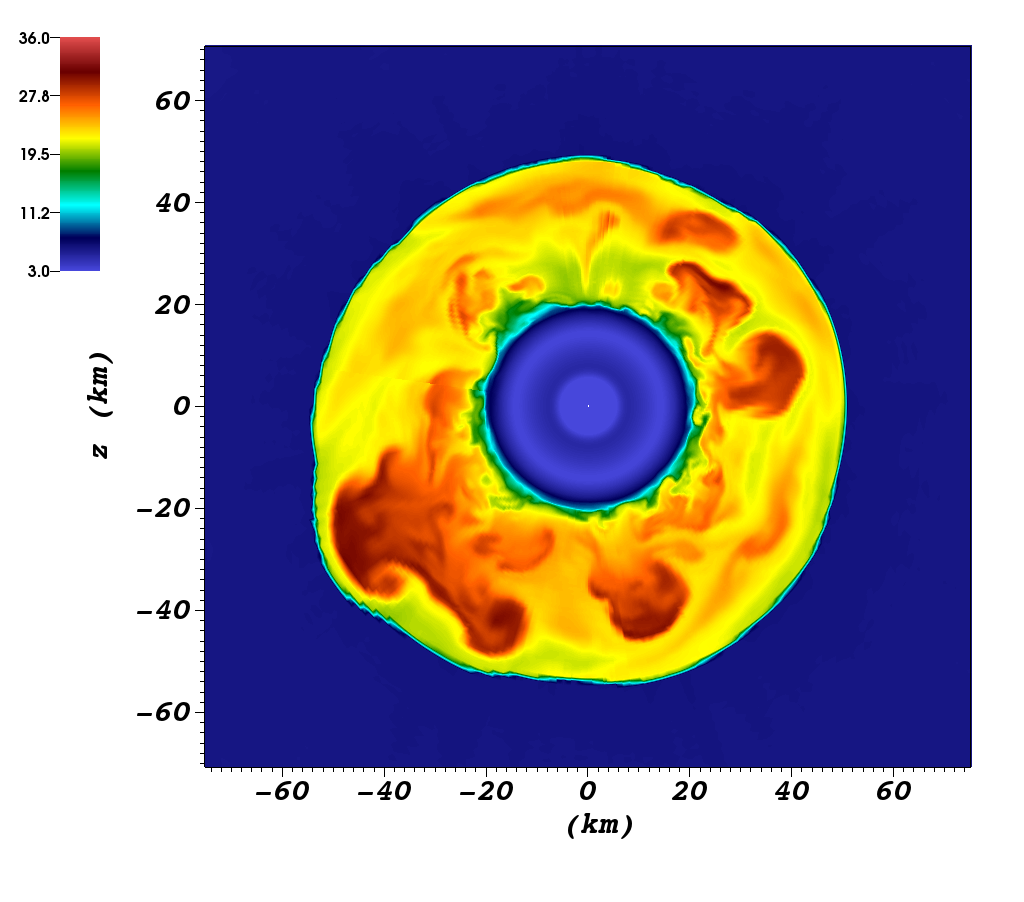}

\includegraphics[width=0.32\textwidth]{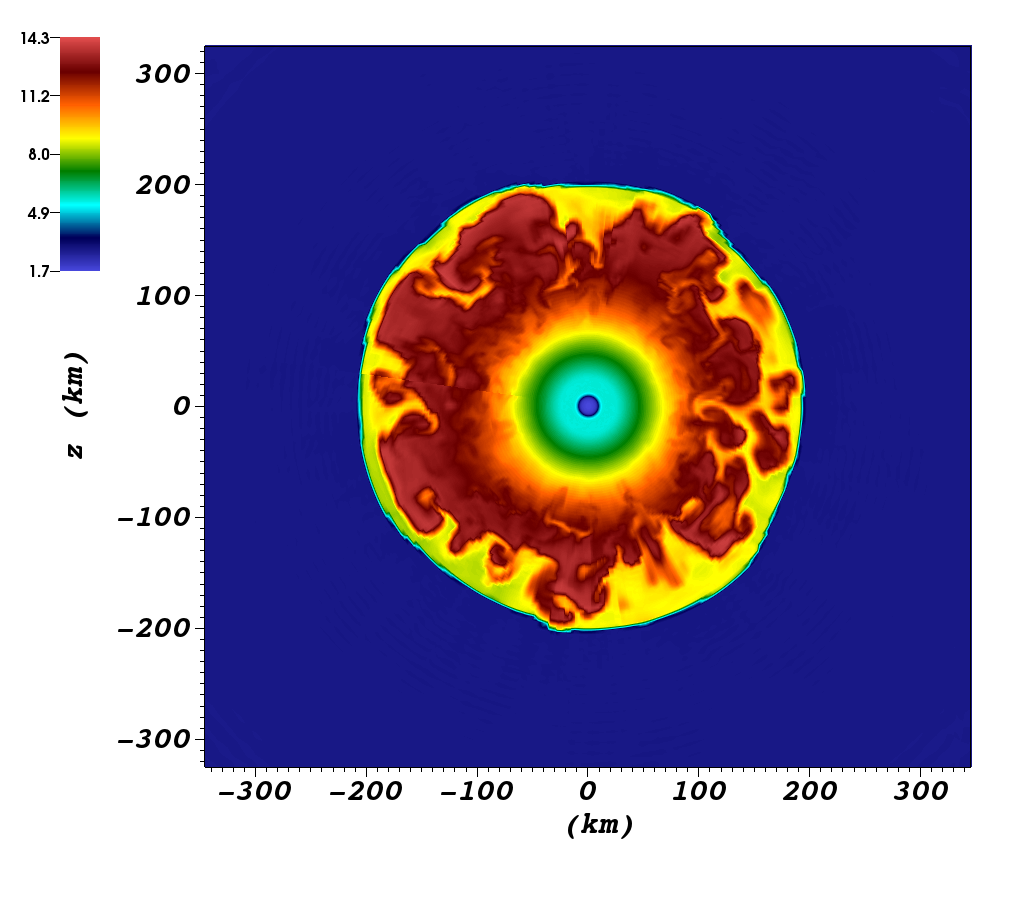}
\includegraphics[width=0.32\textwidth]{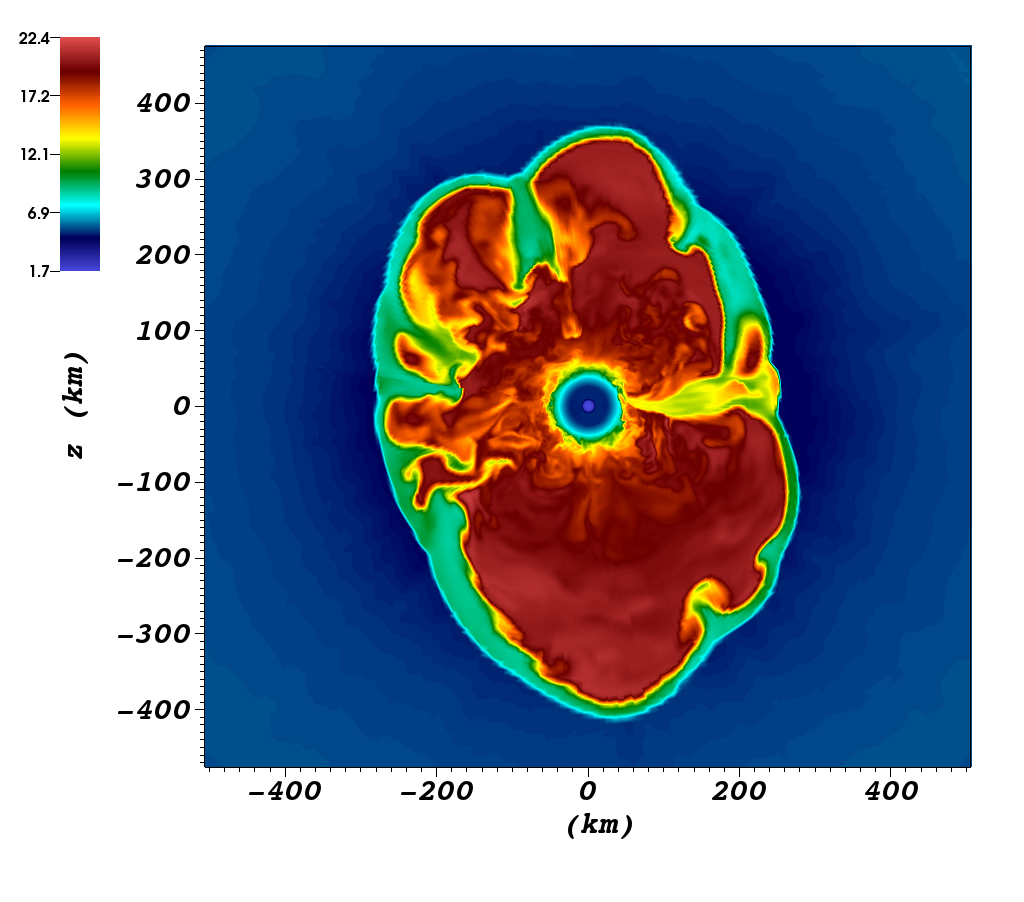}
\includegraphics[width=0.32\textwidth]{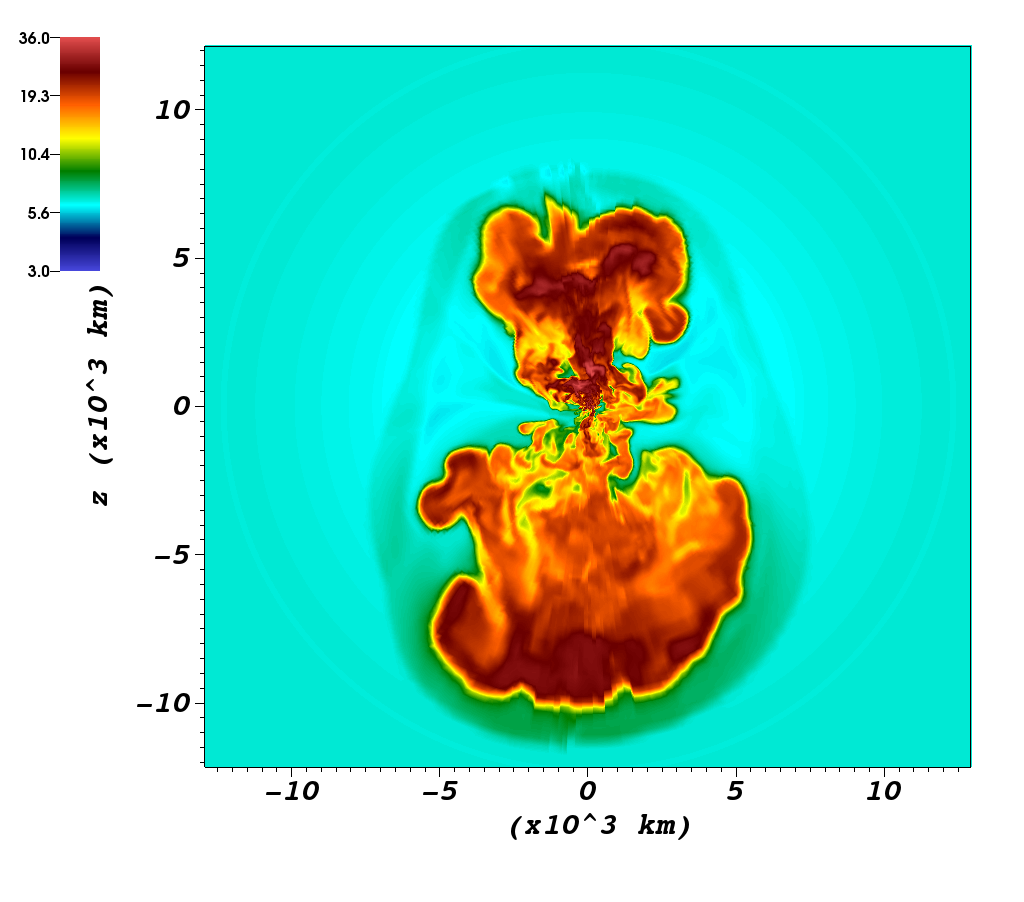}
\caption{From top to bottom we show 2D slices of the entropy for models m39, s18np and y20. For model m39 we show the entropy at 72\,ms, 170\,ms, and 956\,ms after bounce. For model s18np we show the entropy at 124\,ms 310\,ms and 514\,ms after bounce. For model y20 we show the entropy at 95\,ms 195\,ms and 1065\,ms after bounce.}
\label{fig:plots}
\end{figure*}

\begin{figure*}
\centering
\includegraphics[width=\columnwidth]{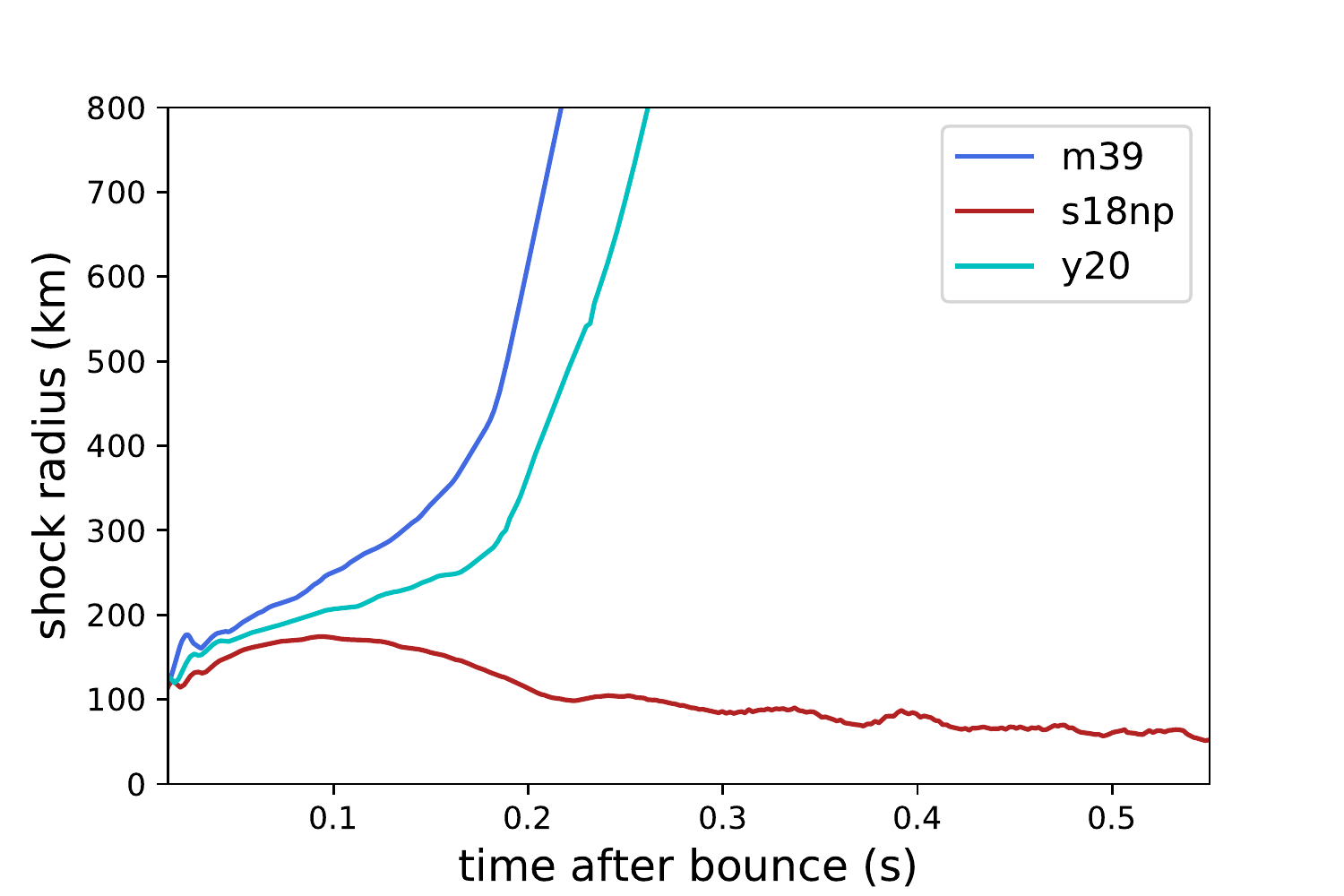}
\includegraphics[width=\columnwidth]{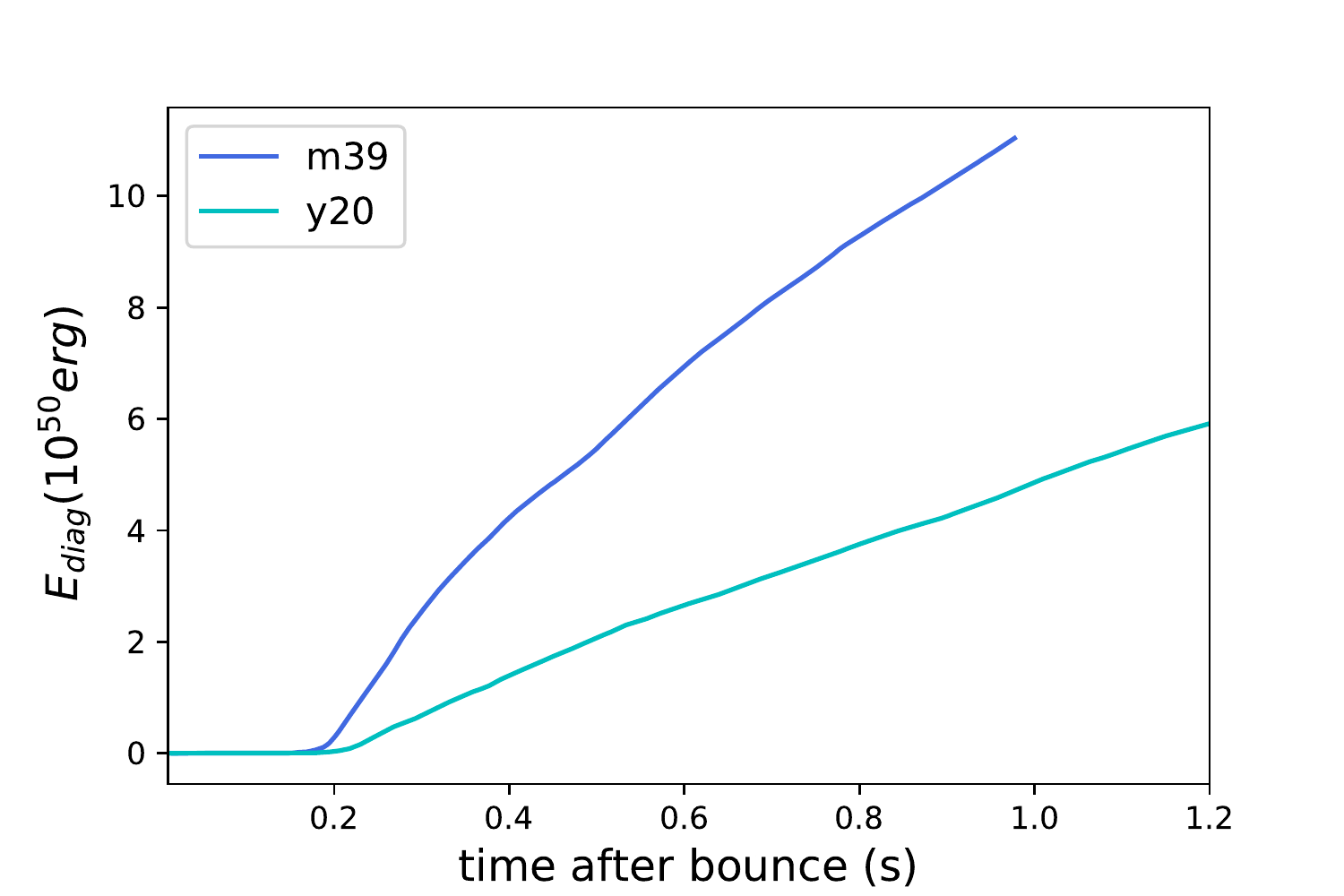}
\caption{(Left) Average shock radius for all simulations. Models m39 and y20 explode about $200\, \mathrm{ms}$ after core bounce. 
Model s18np had not achieved shock revival by the end of the simulation.
(Right) The explosion energy of m39 and y20. Both models have not yet reached their final explosion energies
by the end of the simulation.}
\label{fig:energies}
\end{figure*}

\begin{figure*}
\centering
\includegraphics[width=\columnwidth]{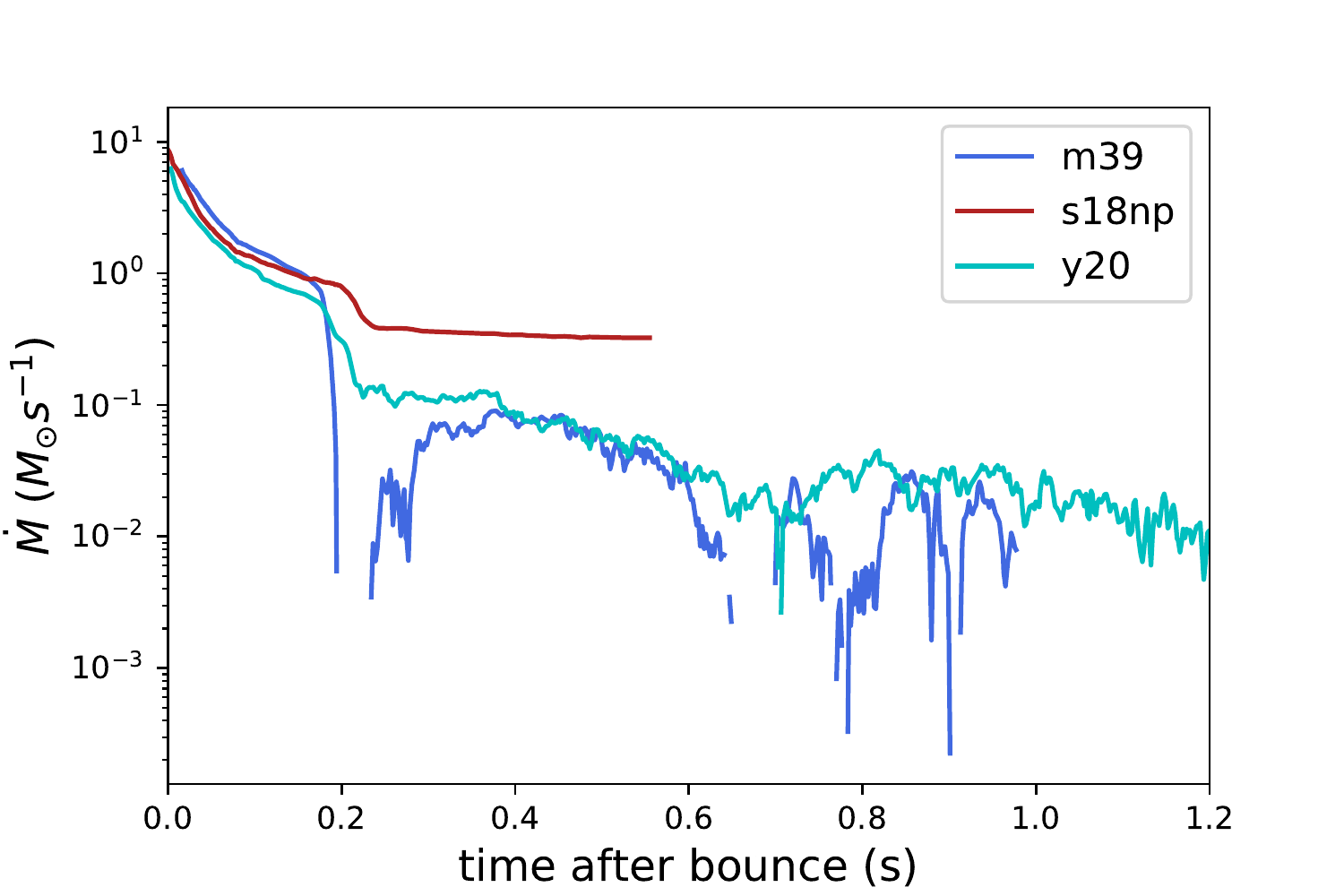}
\includegraphics[width=\columnwidth]{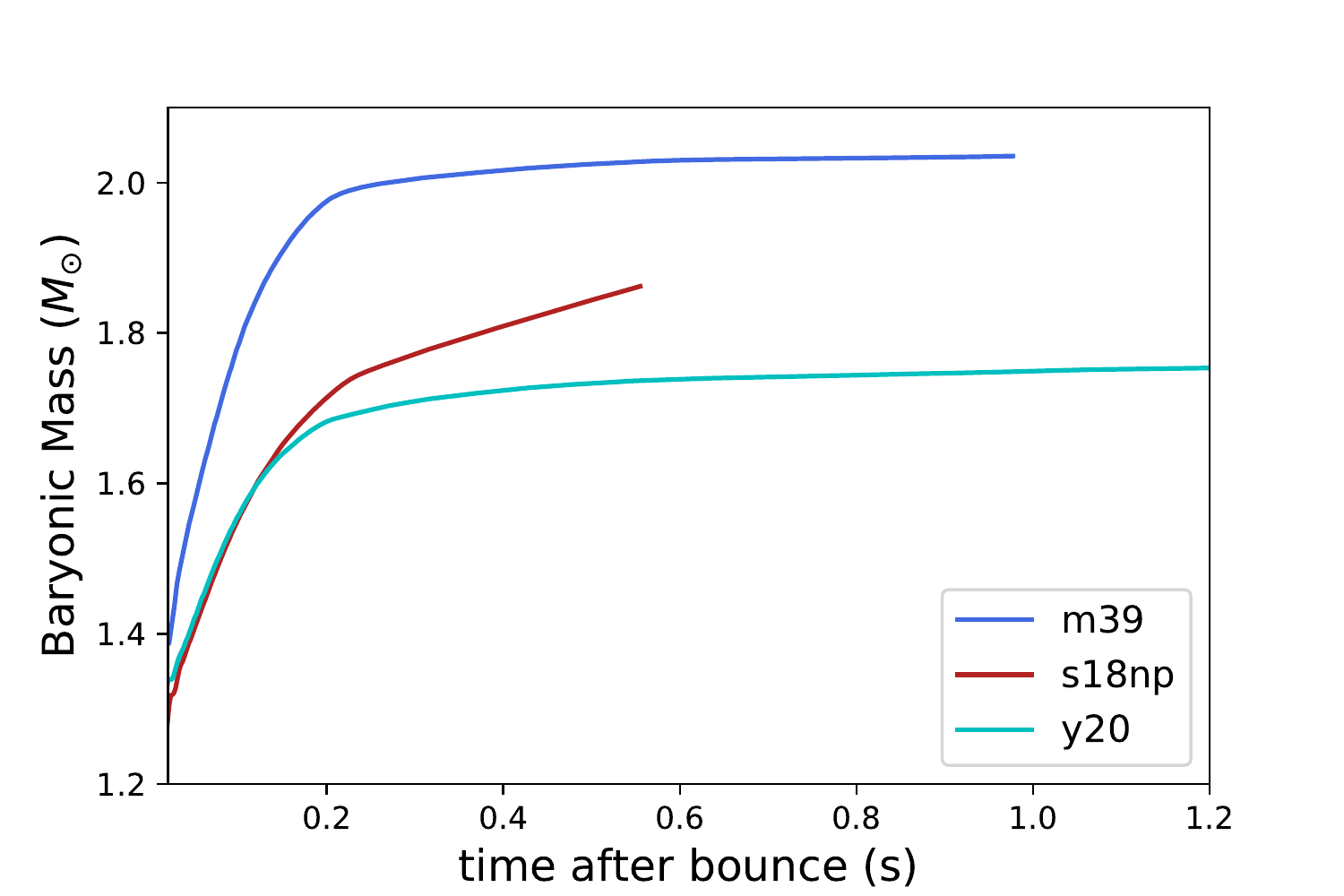}
\caption{(Left) The mass accretion rate. 
Models m39 and y20 experience a sharp drop in mass accretion when the shock is revived, 
but continue to accrete some mass until the end of the simulation. Model s18np has a high 
mass accretion rate up to the end of the simulation. 
(Right) Baryonic mass of the PNS. The PNS mass of model s18np is still increasing due to 
the high accretion rate. Models y20 and m39 have asymptoted to their final values. }
\label{fig:masses}
\end{figure*}

The evolution of the three models is illustrated by meridional 2D
slices at selected times (Figure~\ref{fig:plots}),
and by plots of
the average shock radius and the explosion energies in Figure \ref{fig:energies}.
Models m39 and y20 both undergo neutrino-driven shock revival about $200 \, \mathrm{ms}$ 
after core bounce. Both models are characterised by a bipolar
explosion geometry.
Model s18np has not undergone shock revival by the end of the 
simulation ($0.56$\,s after bounce), and shows
long-standing activity of the SASI spiral mode. The large accretion shock radii
at early post-bounce times
and the precipitous explosions of the
two massive Wolf-Rayet progenitors warrants some critical discussion about the
potential role of our transport approximation and other physical uncertainties
that may affect the simulations.
The fate of models m39 and y20 is reminiscent of other massive models in other recent works.
\citep{2018ApJ...855L...3O,2019arXiv190904152B,burrows_19b}, in which the 
accretion shock radius expands continuously in the post-bounce phase and reaches
large radii of $200 \, \mathrm{km}$ and above early on. This suggests
that such early shock revival as seen for m39 and y20 is plausible. 
A possible explanation for the unusually large early shock radii
comes from the time dependence of the mass accretion rate $\dot{M}$
(Figure~\ref{fig:masses}), and points to structural differences in
the progenitor core between the \textsc{MESA} 
progenitor models y20 and m39 on the one hand,
and the \textsc{Kepler} model  s18np on the other hand.
In model s18np, the decline of the mass accretion rate slows down
more recognisably from about $80 \, \mathrm{ms}$ onward with a
visible kink in the curve, whereas the decline of $\dot{M}$ 
slows less appreciably in y20 and m39. It is during this phase
that model s18np diverges decisively from the other two.

However, the shock radius in models y20 and m39 is already
larger immediately after bounce. This may be connected
to to a deficiency of the deleptonisation scheme of
\citet{liebendoerfer} that we use during the collapse phase,
which tends to overestimate the mass of the homologous
inner core for higher progenitor core entropies
\citep{2008PhRvD..78f4056D}, and thus produces an overly
energetic bounce. Furthermore, our models may overestimate
the PNS and accretion shock radius somewhat because we
neglect electron neutrino pair annihilation as a source
for heavy flavour neutrinos \citep{buras_03} and hence
underestimate the cooling of the neutron star mantle.
On the other hand, models m39 and y20 do not include convective
seed perturbations in the progenitor or magnetic fields, both
of which could be beneficial for early shock revival.
Clearly, more work is needed to reliably determine the fate of
progenitors with such massive helium cores. However,
the present models \emph{can} tentatively teach us about
the expected explosion and compact remnant properties
and the gravitational-wave signals from these massive
progenitors \emph{provided} that they explode reasonably
early.

The models did not reach their final explosion energies
before the end of the simulation, however model m39 has already reached over $10^{51}$\,erg, and
model y20 has reached $6\times10^{50}$\,erg, which are much larger values than those found in 
most other modern 3D simulations of neutrino-driven explosions.
This is encouraging and demonstrates that neutrino-driven explosions
need not be underenergetic, and can reach typical explosion energies
of normal hydrogen-rich and stripped-envelope supernovae
\citep{kasen_09,pejcha_15b,taddia_18}.  With a growth rate of $10^{51}\, \mathrm{erg}\, \mathrm{s}^{-1}$,
the final explosion energy of model m39 may well be
considerably higher than the canonical value of $10^{51}\, \mathrm{erg}$.

Combined with similar findings by other groups, who also observed
shock revival in progenitors with massive cores 
in 3D simulations \citep{2018ApJ...855L...3O,kuroda_18}, this
raises broader questions about the nature of hypernova explosions.
Considering efficiency limitations of the neutrino-driven mechanism
\citep{2012ARNPS..62..407J,2016ApJ...821...38S,2016MNRAS.460..742M},
it is unlikely that neutrino heating alone is sufficient to achieve
hypernova energies, which are of order $\sim10^{52}$\,erg
\citep{1998Natur.395..672I, Soderberg:2006vh}.  However, while the
favoured explanation for hypernova explosions in the magnetorotational
paradigm \citep{winteler_12,moesta,2019arXiv190901105O} relies on magnetic fields from the outset, it may
still be possible that neutrino heating plays a dominant role during
the early phase of hypernova explosions by triggering shock revival
and already delivering part of the explosion energy (which is along
the lines of a hybrid regime between neutrino-driven and MHD-driven
explosions suggested by \citealt{2007ApJ...664..416B}).  Magnetic
field effects could take over later to deliver the bulk of the
explosion energy and launch relativistic jets either from a
millisecond magnetar \citep{uzov_92} or from a collapsar disk
\citep{macfadyen}.  In future, models like m39 need to be continued
for a longer duration to determine to what extent a neutrino-driven
``precursor'' explosion can already contribute to hypernova energies.

\begin{figure}
\centering
\includegraphics[width=\columnwidth]{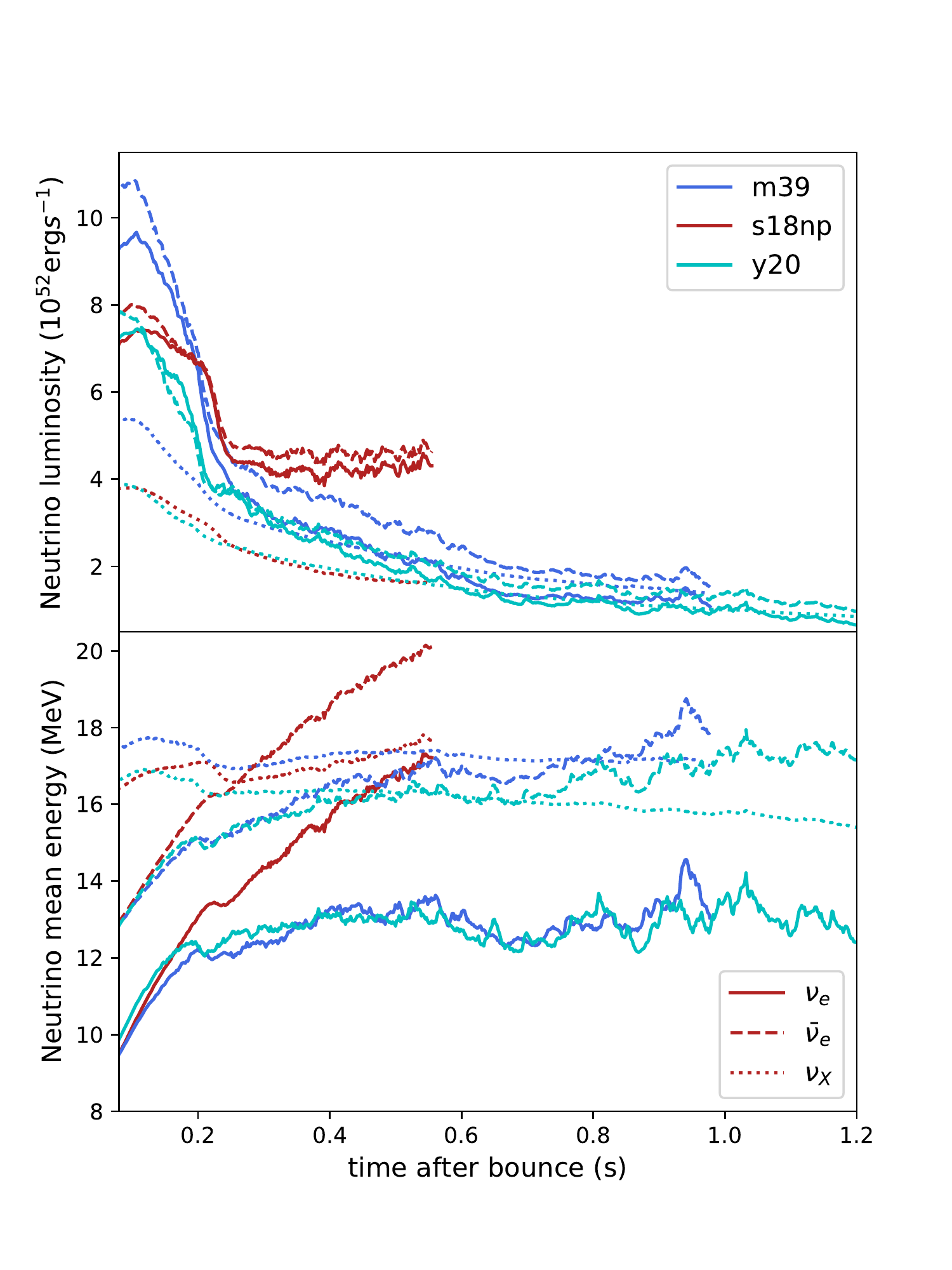}
\caption{Neutrino luminosities (top) and mean energies (bottom) of $\nu_{e}$ (solid), 
$\bar{\nu}_{e}$ (dashed), and heavy flavour neutrinos $\nu_{X}$ (dotted) for models 
m39, s18np and y20.  }
\label{fig:neutrino}
\end{figure}

For the sake of completeness, we also show the neutrino luminosities
and mean energies in Figure \ref{fig:neutrino}.  In the first 200\,ms,
model m39, which has by far the biggest core among
the three models, reaches very high neutrino luminosities. After the 
onset of the explosion
in models m39 and y20, model s18np has much larger neutrino
luminosities and neutrino energies than the other models.
The switch in the hierarchy of the models (with model
s18np overtaking model m39 in terms of
luminosity and mean energies) is a natural consequence
of the different mass accretion rates.
 All the models show the characteristic drop in the mass accretion
rate that is associated with the infall of the Si/O layer. For models
y20 and m39, this drop triggers shock revival, and they
continue to accrete some mass only at a low level until the end of the
simulation.  The mass accretion rate for model s18np remains high
towards the end of the simulation.

\section{Remnant Properties}
\label{sec:properties}

\begin{figure*}
\centering
\includegraphics[width=\columnwidth]{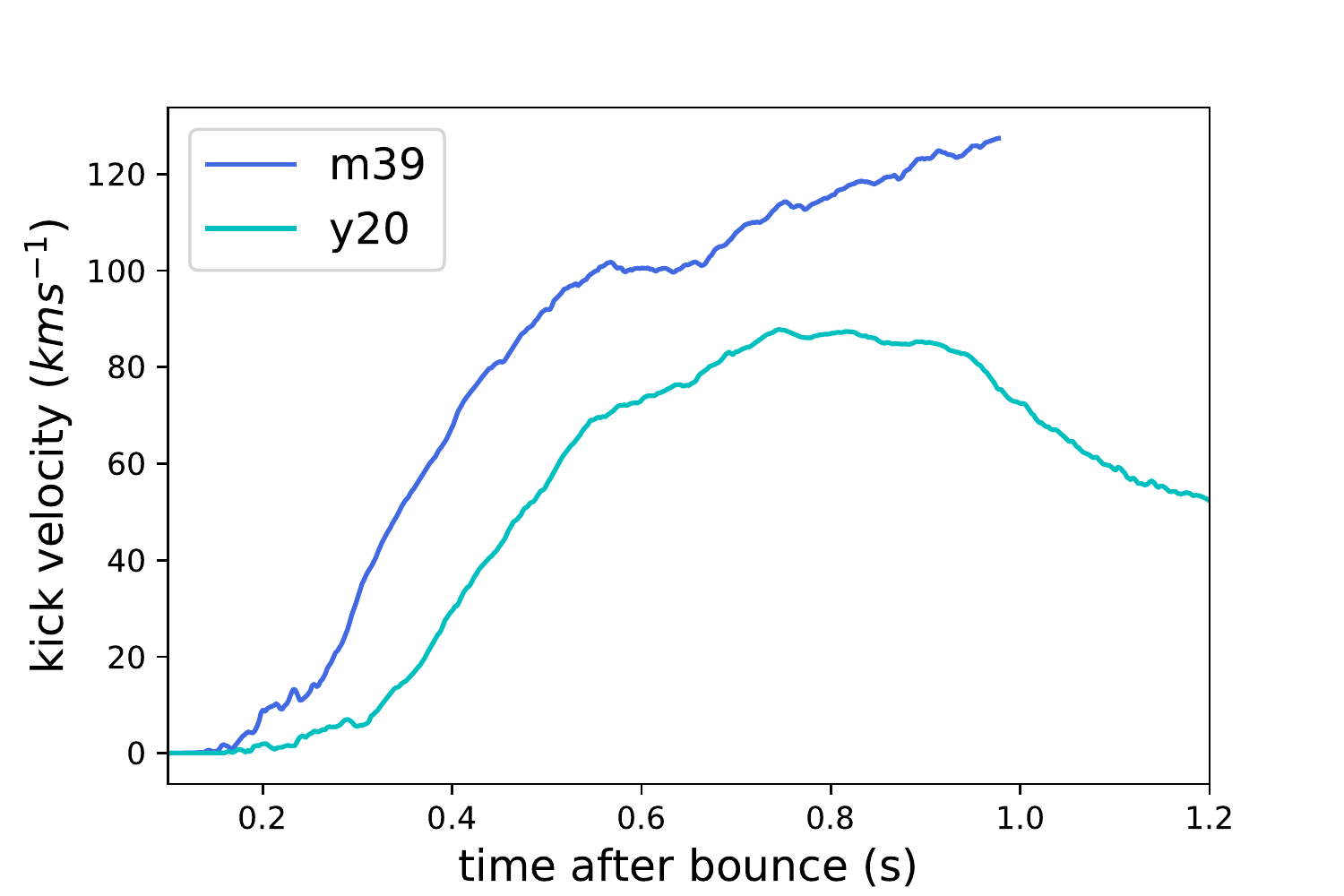}
\includegraphics[width=\columnwidth]{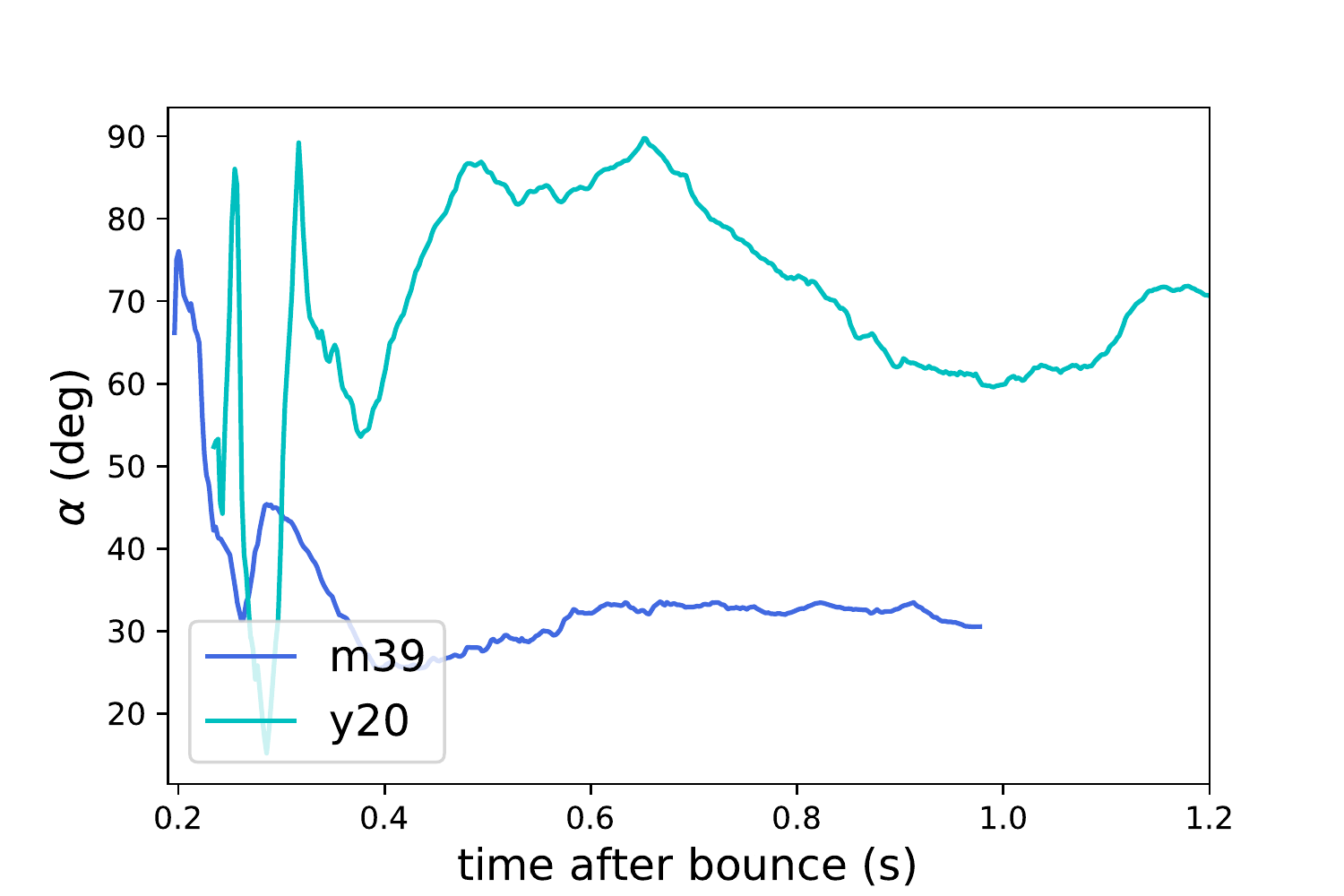}
\caption{
(Left) The kick velocities of models m39 and y20. They do 
not reach their final values before the end of the simulation.
(Right) Evolution of the angle between the spin and kick direction. 
We plot the spin-kick angle only once the explosion energy reaches a significant 
positive value and the spin and kick direction do not vary rapidly
any more. }
\label{fig:spin_kick_angle}
\end{figure*}

The baryonic PNS masses for all models are shown in Figure~\ref{fig:masses}
and Table~\ref{tab:properties},
(although the PNS mass for the non-exploding
model s18np is of less interest for comparing with the observed
neutron star population). The exploding models m39
and y20 have almost asymptoted to their final mass values
of $2.04\,M_{\odot}$ and $1.75\,M_{\odot}$ respectively,
provided that there is no late time fallback. 
Their masses are within the range expected from observations
\citep{2016ARA&A..54..401O, 2019ApJ...876...18F}. 

In Figure~\ref{fig:spin_kick_angle}, we show the neutron star kicks
and the angle $\alpha$ between the spin and kick direction. For evaluating
the kick and spin, we followed the procedure of \citet{wongwathanarat,2019MNRAS.484.3307M}.

 The  kick velocities for
both model m39 and y20 have not reached their final values before the
end of the simulation times.  The kick velocities at the end of
the simulations are within the observed distributions of neutron star
kicks \citep{2005MNRAS.360..974H, 2006ApJ...643..332F,
  2007ApJ...660.1357N}.  In fact, the kick velocities are rather on
the low side of the observed mean value of $\mathord{\sim} \, 265 \, \mathrm{km}\,\mathrm{s}^{-1}$
\citep{2005MNRAS.360..974H, 2006ApJ...643..332F, 2007ApJ...660.1357N}.  
Model m39 has only reached a kick velocity of
$127\,\mathrm{kms}^{-1}$ by the end of the simulation, and the kick
for y20 is even smaller. These rather lower kick velocities are the
result of the bipolar explosion geometry of both models
(Figure~\ref{fig:plots}). The low kick velocities have implications
for the proposed correlation between explosion energy and kick
velocity \citep{10.1093/mnras/stw1275,janka,10.1093/mnras/sty2463}. Our results constitute
further evidence that this correlation is not a tight one.  More
likely, the explosion energy only determines the maximum
\emph{attainable} kick velocity, below which one expects a distribution
of kicks depending on the degree of unipolarity or bipolarity
\citep{2019MNRAS.484.3307M}. 

The evolution of the spin-kick angle $\alpha$ in model m39 is
noteworthy.  Observations have suggested that the birth spins and
kicks of neutron stars tend to be aligned \citep{2005MNRAS.364.1397J,
  2007ApJ...660.1357N, 2013MNRAS.430.2281N}.  However, even though
hydrodynamic mechanisms for spin-kick alignment have been proposed
\citep{janka_17}, no trend towards spin-kick alignment has been observed so
far in 3D CCSN simulations of non-rotating progenitors
\citep{wongwathanarat,2019MNRAS.484.3307M}. For our non-rotating
explosion model y20, we also find a large angle $\alpha\approx
70^\circ$ between the spin and kick. For the rapidly rotating model
m39, however, the spin-kick angle stays at about $\alpha \approx
30^{\circ}$. In this case, spin-kick
alignment comes about because the bipolar outflows are aligned
with the initial rotation axis, which is not changed appreciably
by asymmetric accretion onto the PNS.
This suggests that progenitor rotation could play
a role in explaining spin-kick alignment (provided that
the observational evidence remains robust).
However, even for model m39, the spin-kick alignment is not perfect,
and is not clear whether the polar alignment of the outflows
is a robust feature for rotating progenitors. In fact, some
previous 3D supernova simulations of rotating progenitors
rather find that the explosion develops preferably
in the equatorial plane \citep{2018MNRAS.475L..91T,summa}. Clearly, more
simulations are required to determine the implications of rotation
for the explosion morphology and spin-kick alignment.

\section{Gravitational Wave Emission}
\label{sec:gw}

\begin{figure*}
\centering
\includegraphics[width=\textwidth]{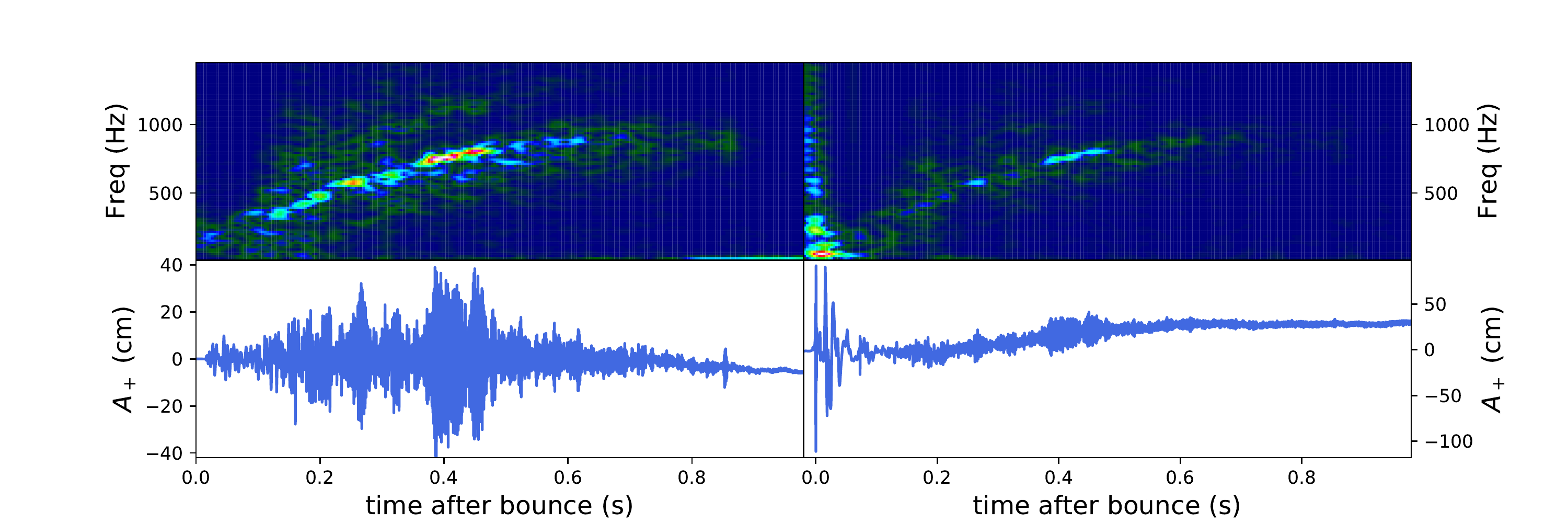}
\includegraphics[width=\textwidth]{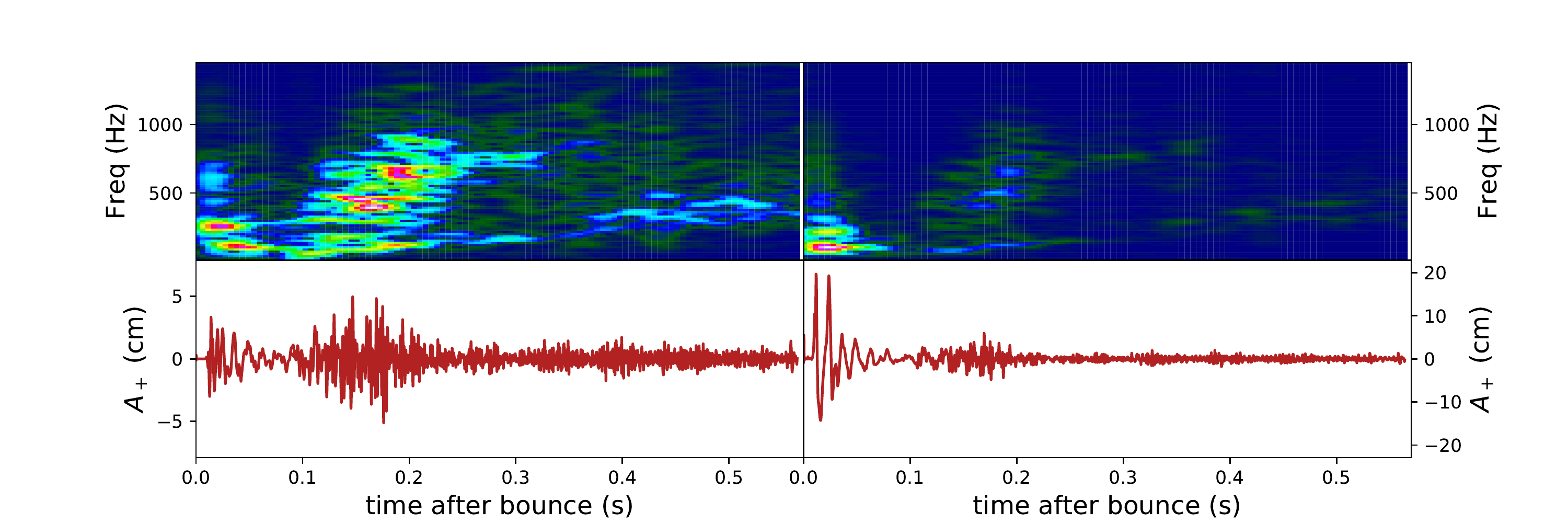}
\includegraphics[width=\textwidth]{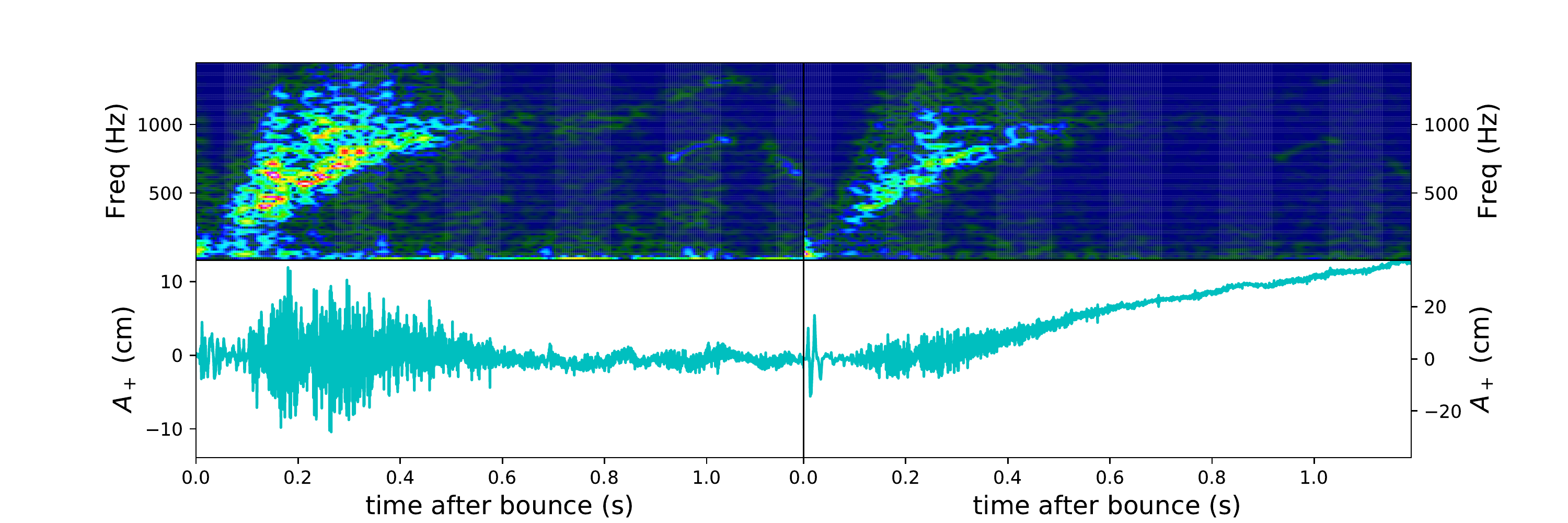}
\caption{The gravitational-wave emission for our models as measured at the pole
(left) and the equator (right). Top is model m39, middle is model s18np, and bottom 
is model y20. All models show the rise of gravitational-wave frequency with time 
associated with the f-mode, and strong prompt convection at the polar observer angles. 
Model s18np also shows low frequency SASI modes. All models
have their highest amplitudes within the first 0.5\,s after bounce. The rotation in
model m39 results in very high amplitude gravitational-wave emission. }
\label{fig:gw_emission}
\end{figure*}

The gravitational-wave emission for our models, as measured by an observer at the 
pole and the equator, is shown in Figure \ref{fig:gw_emission}.
The high mass and rotation of model m39 results in high gravitational-wave amplitudes 
of up to 40\,cm at the pole and over 100\,cm at the equator.
Model y20 reaches amplitudes of $\sim10$\,cm at the pole,
and $\sim18$\,cm at the equator. Model s18np has a much lower amplitude of $\sim5$\,cm at 
the equator, due to the lack of shock revival in this model, however its prompt 
convection signal in the direction of the equator reaches amplitudes as high as model y20.

\begin{figure*}
\centering
\includegraphics[width=0.62\textwidth,height=8cm]{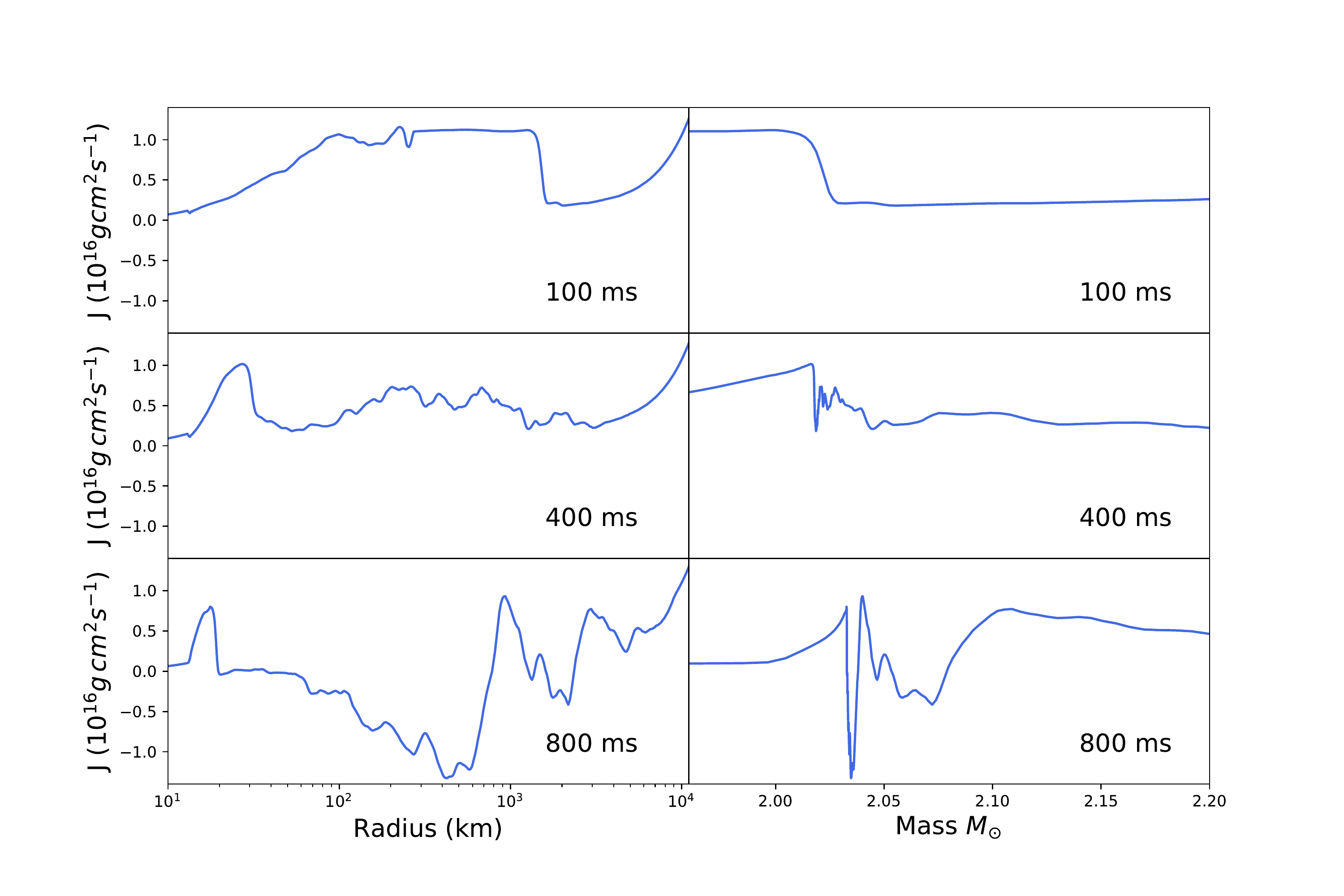}
\includegraphics[width=0.32\textwidth,height=8cm]{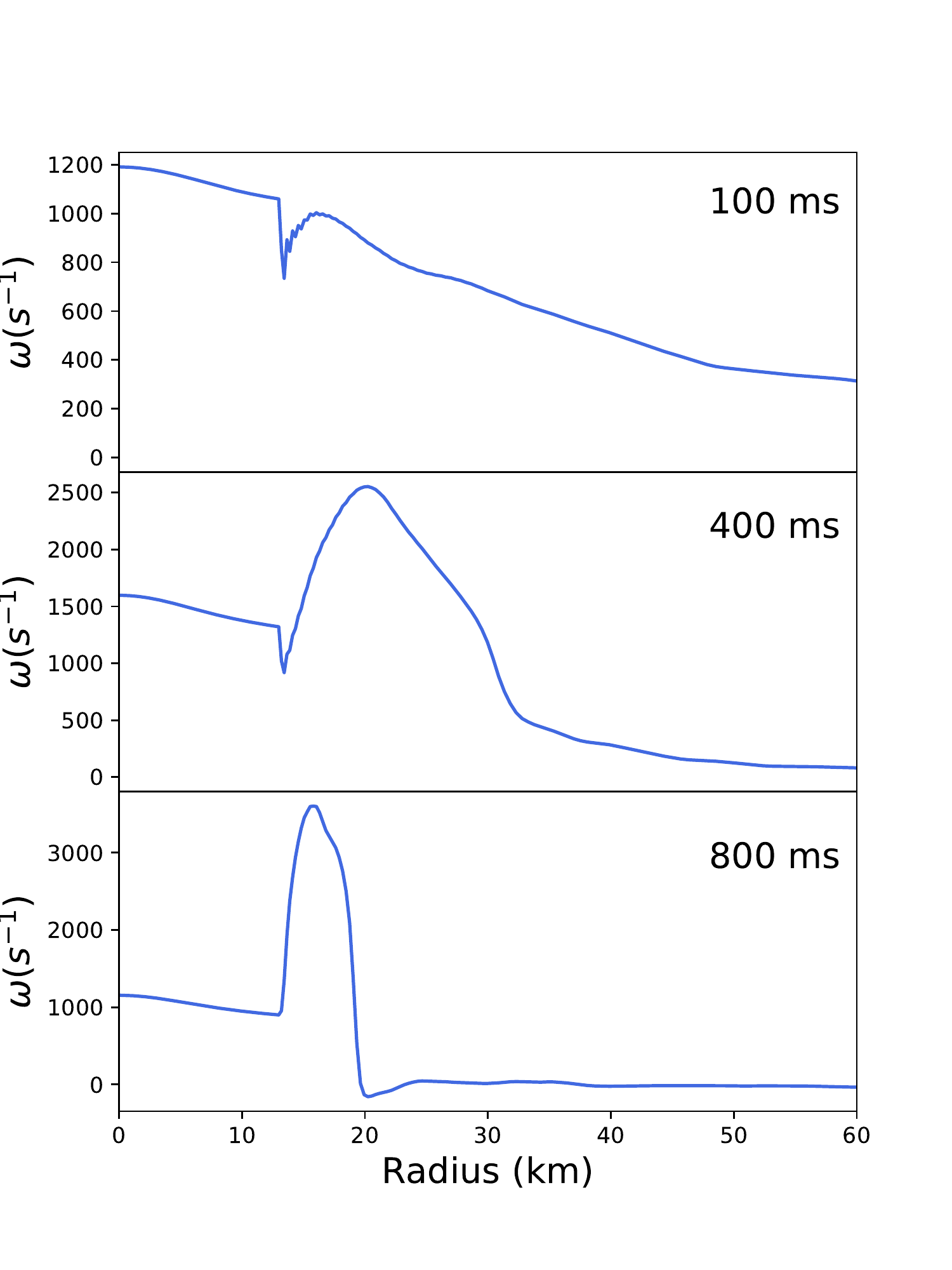}
\caption{(Left) The angular momentum of model m39 as a function of radius and mass 
coordinate at 100\,ms, 400\,ms and 800\,ms after bounce. (Right) The angular velocity 
of model m39 as a function of radius at 100\,ms, 400\,ms and 800\,ms after bounce. }
\label{fig:angular}
\end{figure*}
 
All of our models show gravitational-wave emission associated with prompt convection.
Prompt convection occurs shortly after bounce time due to an unstable negative entropy 
gradient created by the weakening of the prompt shock and 
large neutrino losses during the neutronisation burst \citep{1994ApJ...433L..45B}.
Models y20 and m39 both have strong prompt convection signals shortly after bounce
that are visible only in observer angles towards the poles. Model s18np shows a 
signal due to prompt convection at all angles. The emission due to prompt convection
occurs only for the first few ms at frequencies below 200\,Hz. In model m39, there is
also power at higher frequencies $> 500\, \mathrm{Hz}$ as expected from a rotational
core bounce \citep{2007PhRvL..98y1101D}. The frequency of this signal
is set by the fundamental $l=2$ mode of the PNS \citep{2015MNRAS.450..414F},
but the power of this components is spread across
a wider range in frequency. This could suggest that either higher-frequency p-modes
are excited during bounce, or may simply be a signal analysis artefact related
to the short duration of the bounce signal and/or to edge effects.
After shock revival, we find a large tail from asymmetric shock propagation
\citep{2009ApJ...707.1173M,2013ApJ...766...43M}
that occurs in the emission of model y20, 
and a smaller tail is observed in model m39. 

The time-frequency structure of all three models show the characteristic emission that is typically 
associated with f-mode oscillations of the PNS
\citep[e.g.][]{2009ApJ...707.1173M,2013ApJ...766...43M,2018ApJ...861...10M}. Previous studies have shown that the 
gravitational-wave frequency of the f-mode is given by,
\begin{equation}
f\approx \frac{1}{2\pi}\frac{GM_\mathrm{by}}{R_\mathrm{pns}^2}
\sqrt{1.1\frac{m_\mathrm{n}}{\langle E_{\bar{\nu}_\mathrm{e}}\rangle}}
\left(1-\frac{GM_\mathrm{by}}{R_\mathrm{pns}c^2}\right)^{2},
\label{eqn:gmode}
\end{equation}
where $M_\mathrm{by}$ is the baryonic mass of the PNS, $R_\mathrm{pns}$ is the
radius of the PNS, $m_\mathrm{n}$ is the neutron mass, and $E_{\bar{\nu}_\mathrm{e}}$ 
is the electron antineutrino mean energy \citep{2013ApJ...766...43M}. We find that
this approximation of the f-mode frequency matches well with our models y20 
and s18np during the high amplitude periods of the gravitational-wave emission.
In model s18np, the f-mode rises in frequency more rapidly than in our other models, 
but its amplitude peaks at a lower frequency of $\sim360$\,Hz because the f-mode
only experiences strong forcing early on before the shock retracts and the mass
and turbulent kinetic energy in the gain region shrink. 
Model y20 also has a lower frequency mode that is clearly visible in the spectrogram
after about $0.9 \, \mathrm{s}$, and decreases slowly in frequency up to the end of the simulation. 
The frequency trajectory for this mode is compatible with a g-mode or hybrid mode
\citep{2018MNRAS.474.5272T,2018ApJ...861...10M}.

The higher f-mode frequency trajectory in model s18np compared to model y20 is easily
understood as a consequence of the steeper rise of the PNS mass in this non-exploding model.
The f-mode frequencies in our rotating model m39 are puzzling at first glance. Due to the
high PNS mass and only modestly high neutrino mean energies (lower than for model s18np),
one would except a very high f-mode frequency.
However, a quantitative analysis shows that Equation~(\ref{eqn:gmode}) no longer provides a good
approximation of the frequency of the gravitational-wave emission for this model. 
The amplitude of the f-mode emission for model m39 is highest at a frequency 
of $\sim750$\,Hz, but Equation~(\ref{eqn:gmode}) predicts higher frequency 
gravitational-wave emission of $\sim 2000$\,Hz 
for this model.

 The reduction of the actual f-mode frequency
in relation to the PNS properties in our rotating model 
can be explained by the angular momentum distribution in the
progenitor (and hence in the PNS).

In Figure \ref{fig:angular}, we show the spherically averaged specific angular momentum $j$ and 
the angular velocity $\omega$ of 
model m39 at 100\,ms, 400\,ms and 800\,ms after bounce. 
The progenitor exhibits a significant drop of $j$ and $\omega$ from
the Fe and Si core to the O shell 
by a factor of several. This translates into a steep negative gradient 
in $j$ and $\omega$ in the PNS surface region after the infall of the Si/O shell interface.
This steep gradient remains stable and even steepens despite the
fact that the PNS surface region loses angular momentum into the
ejecta region due to shear instabilities.
\footnote{In addition, angular momentum conservation is imperfect
in the core region inside a radius of $\mathord{\sim} 10\, \mathrm{km}$
that we simulate in spherical symmetry, the core region
spins down significantly as can be seen from Fig.~\ref{fig:angular}.
This may also indirectly lead to a deceleration of the rotation
in the PNS surface region, i.e.\ our model may actually
\emph{underestimate} the angular momentum gradient at the
PNS surface.
}

Such a negative angular momentum gradient counteracts the stabilisation
of the PNS surface region by the positive entropy gradient. The
oscillation  frequency $N$ of a displaced bubble is modified
according to the Solberg-H{\o}iland criterion,
\begin{equation}
N^2=N_\mathrm{BV}^2+
\frac{1}{\varpi^3}
\frac{\pd j^2}{\pd \varpi}
\sin\theta,
\end{equation}
where $N_\mathrm{BV}=2\pi f$
is the Brunt-V\"ais\"al\"a frequency, and $\varpi=r \sin \theta$
is the distance from the rotation axis. The actual reduction of the
f-mode frequency that dominates the gravitational-wave emission can only
be inferred from a full eigenmode analysis, but since we clearly
find that the Rayleigh discriminant
$N_\omega=(\varpi^{-3} \pd j^2/\pd \varpi)^{1/2}$ reaches values
comparable to $N_\mathrm{BV}$, rotational destabilisation
will reduce the mode eigenfrequency considerably, which explains
the unexpectedly low frequencies in the spectrogram
of model m39.

This has important implications for the interpretation of
gravitational-wave spectrograms. It has been suggested that universal
relations for the mode frequencies akin to Equation~(\ref{eqn:gmode})
can be used to constrain PNS parameters \citep{2019PhRvL.123e1102T}
using the spectrogram for non-rotating progenitors. Since
Equation~(\ref{eqn:gmode}) no longer holds for rapidly rotating
progenitors, it will be considerably more difficult to constrain the
PNS mass, radius, and surface temperature using measured mode
frequencies in this case.

Model s18np shows lower frequency gravitational-wave emission due to
SASI. The gravitational-wave frequency of the SASI roughly follows
the SASI frequency, which is well described
by
\begin{equation}
f_\mathrm{sasi} = \frac{1}{19\,\mathrm{ms}} \left(\frac{R_\mathrm{sh}}{100\,\mathrm{km}}\right)^{-3/2} 
\ln{\left(\frac{R_\mathrm{sh}}{R_\mathrm{pns}}\right)}^{-1}
\end{equation} 
where $R_\mathrm{sh}$ is the shock radius and $R_\mathrm{pns}$ is the radius of the PNS \citep{2014ApJ...788...82M}.
The frequency is in the most sensitive range for ground based gravitational-wave detectors. 
The SASI features are present in the gravitational-wave signal for a longer duration 
than in some other recent 3D models \citep{2016ApJ...829L..14K,2017MNRAS.468.2032A,2018arXiv181007638A}, due to 
our longer simulation time and the lack of shock revival in this model.  
As a consequence, the frequency of the SASI signal reaches
up to $\mathord{\sim} 400 \, \mathrm{Hz}$.

\subsection{Detection prospects}
\label{subsec:detect}

\begin{figure*}
\centering
\includegraphics[width=0.45\textwidth]{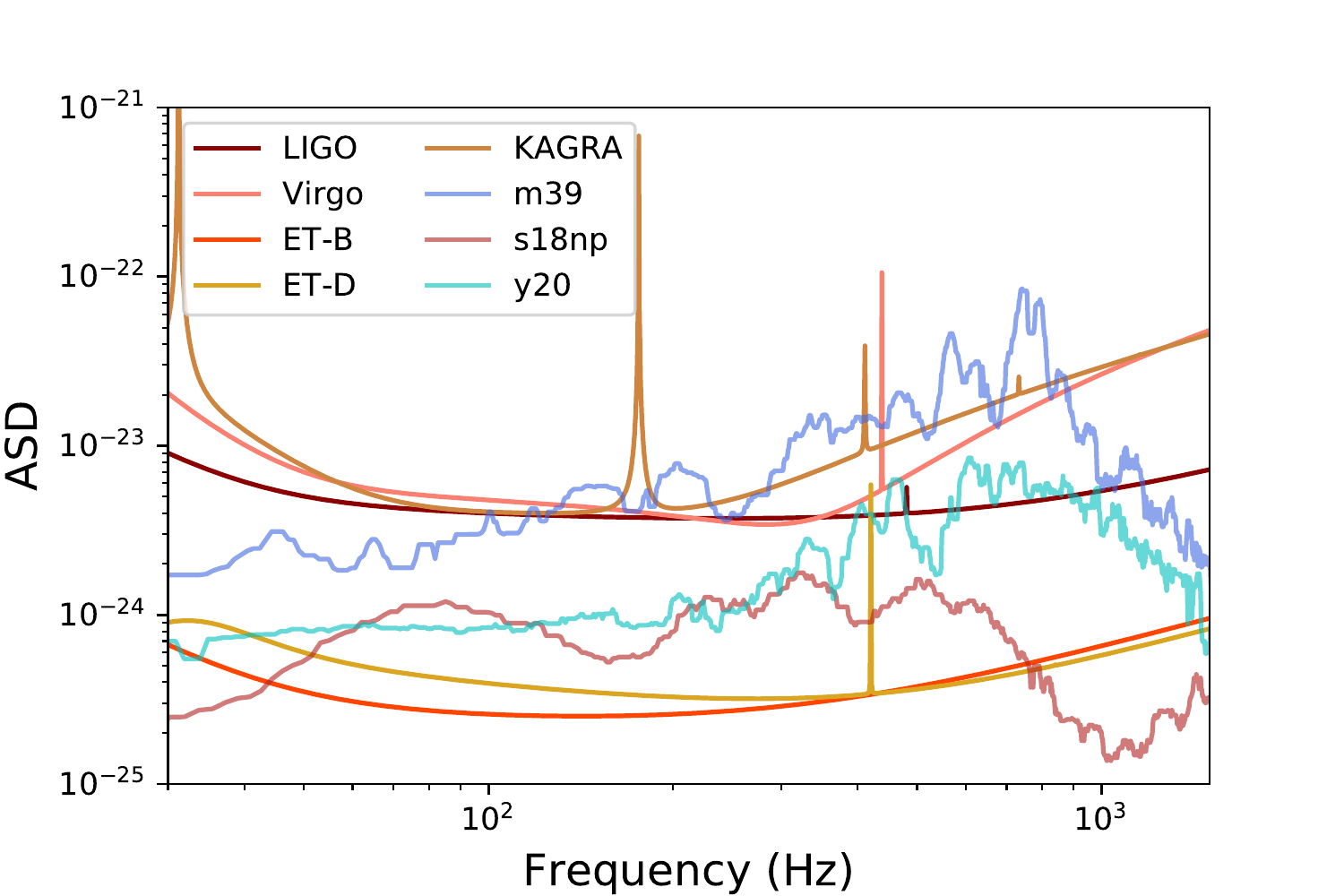}
\includegraphics[width=0.45\textwidth]{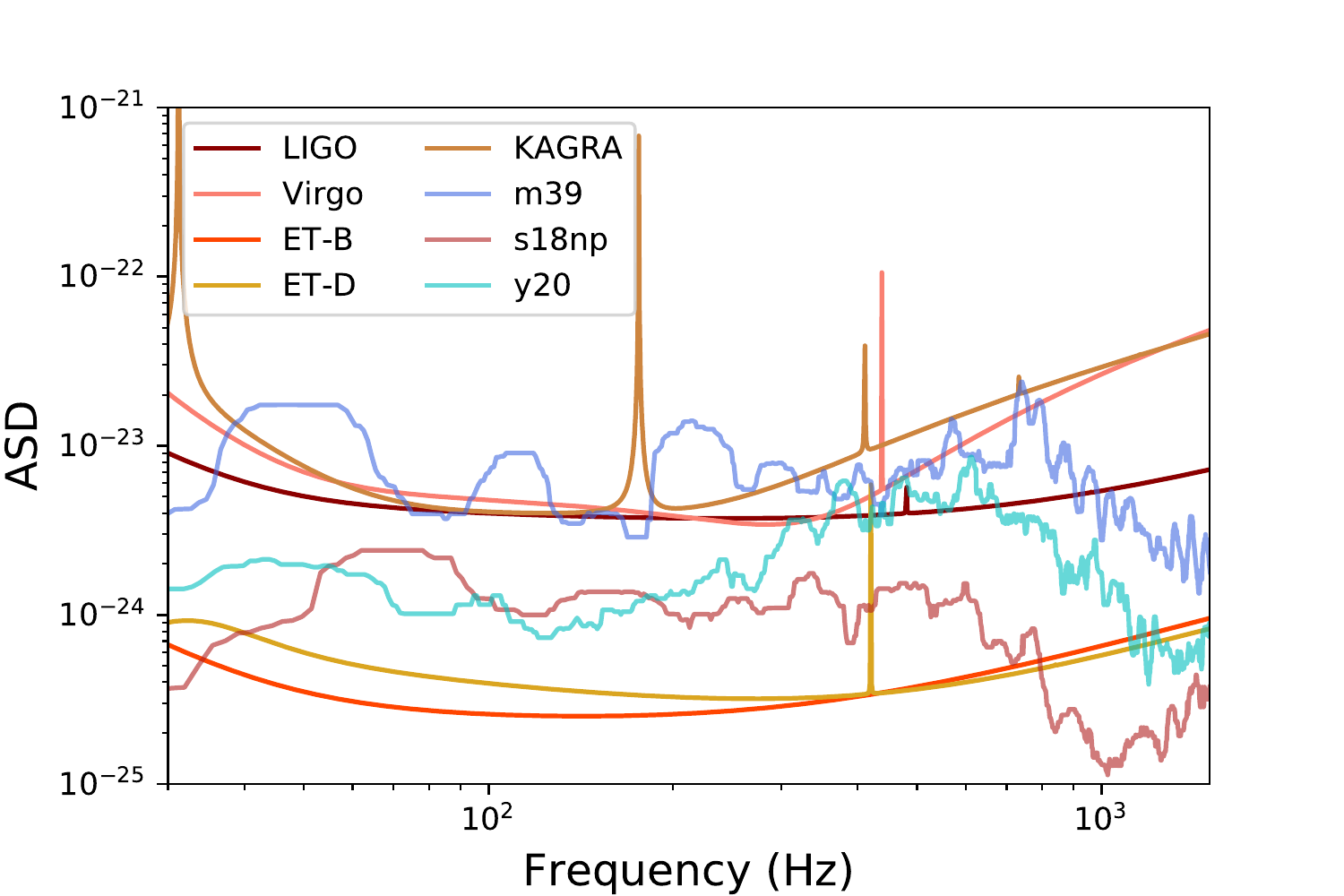}
\caption{The amplitude spectral density (ASD) of our three models at a distance of 100\,kpc, 
and the predicted design sensitivity noise curves of current and future ground-based 
gravitational-wave detectors. The left and right panel show the
ASD for an observer in the polar direction and the equatorial plane.
}
\label{fig:gw_asd}
\end{figure*}

In Figure \ref{fig:gw_asd}, we show the amplitude spectral density (ASD) of our models 
as measured at the pole and equator at a 
distance of 100\,kpc, and the ASD noise curves for current and future ground-based
gravitational-wave detectors. We include design sensitivity curves for aLIGO and AdVirgo, the 
Japanese underground detector KAGRA \citep{2018LRR....21....3A}, 
and two different proposed designs for ET referred to as ET-B \citep{2008arXiv0810.0604H}
and ET-D \citep{Hild_2011}. Model s18np has gravitational-wave emission across the entire
frequency bands of the detectors. Model y20 peaks at higher frequencies where KAGRA
and Virgo are less sensitive. Model m39 has stronger high frequency emission at the 
pole, and a stronger lower frequency signal at the equator due to the high amplitude
signal from the rotational bounce and prompt convection. 

We calculate the maximum distance out to which the gravitational-wave emission from our models
can be detected in current and future gravitational-wave detectors. We assume the threshold for 
detection is given by an optimal matched filter signal to noise ratio (SNR) of 8, where the 
optimal SNR is given by
\begin{equation}
\mathrm{SNR} = \sqrt{ 4  \sum_i  \frac{h(f_i)^{2}}{\sigma(f_i)} \Delta f }
\end{equation}
where $\Delta f$ is the frequency spacing, $\sigma(f_i)$ is the power spectral density
of the detector noise at frequency $f_i$, 
and $h(f_i)$ is the gravitational-wave signal in the frequency domain
at frequency $f_i$ \citep{1994PhRvD..49.2658C}.

 The results for 
our models, using the gravitational-wave emission at both the equator and pole 
observer directions, are shown in Table \ref{tab:distance}. All models have higher 
maximum detectable distances in the aLIGO detector than in AdVirgo and
KAGRA due to the better high frequency sensitivity of the aLIGO detector. In current 
detectors, model s18np and y20 are detectable only in our Galaxy. Model m39 reaches a 
few hundred kpc, beyond the Large Magellanic Cloud at 50\,kpc, and M31 at 77\,kpc. 
The detection distances for m39 are comparable to those of the magnetorotational 
models\footnote{Some of these gravitational wave forms came from
non-magnetised models with rapid rotation that would presumably
result in magnetorotational explosions if efficient magnetic
field amplification processes operate during the post-bounce phase.} in \citet{2016PhRvD..93d2002G}. Interestingly, the large detection distances
for m39 are only modestly direction-dependent and do not rest very much on the
rotational bounce signal. The detection distance varies by less than a factor
of two between the polar and equatorial direction.
For an observer in the equatorial plane, much of the SNR
comes from the low-frequency emission due to prompt convection, while
the long-duration high-frequency emission is the most important contribution
in the polar direction.
In the future gravitational-wave detector ET, model m39 is detectable out to almost 
2\,Mpc (i.e., through the Local Group), and models y20 and s18np can be detected up to a few hundred kpc. ET-B has 
better lower frequency sensitivity than ET-D, which leads to large increase in 
detectable distance for SASI-dominated signals like model s18np.

\begin{table*}
\begin{center}
\begin{tabular}{|c|c|c|c|c|c|c|} 
 \hline
 & m39 pole & m39 eqt & s18np pole & s18np eqt & y20 pole & y20 eqt\\ 
 \hline
LIGO & 190\,kpc & 119\,kpc & 14\,kpc & 22\,kpc & 32\,kpc & 32\,kpc \\ 
 \hline
Virgo & 90\,kpc & 89\,kpc & 12\,kpc & 20\,kpc & 18\,kpc & 20\,kpc \\
 \hline
KAGRA & 60\,kpc & 77\,kpc & 11\,kpc & 18\,kpc & 13\,kpc & 18\,kpc \\
 \hline
ET-B & 1.9\,Mpc & 1.55\,Mpc & 190\,kpc & 320\,kpc & 330\,kpc & 370\,kpc \\
 \hline
ET-D & 1.9\,Mpc & 1.21\,Mpc & 155\,kpc & 240\,kpc & 330\,kpc & 340\,kpc \\
 \hline
 \end{tabular}
\caption{The distance needed for our models to achieve an optimal signal to noise ratio of 8 
in current and future ground based gravitational-wave detectors. The better high-frequency
sensitivity of aLIGO, where the f-mode emission has the strongest amplitudes, results in 
larger detection distances than AdVirgo and KAGRA. ET-B is more sensitive to SASI emission 
than ET-D. 
}
\label{tab:distance}
\end{center}
\end{table*}

It is interesting to put these numbers into perspective by comparing
them to the estimated detection distances for some other recent 3D
supernova models, even though these distances have sometimes been determined in
a somewhat different way. Maximum detection distances $ \mathord{\sim
  15 \, \mathrm{kpc}}$ with aLIGO and $ \mathord{150\texttt{-}300\ \,
  \mathrm{kpc}}$ with 3G instruments have been reported for most
non-rotating 3D models
\citep{2017MNRAS.468.2032A,2019PhRvD.100d3026S,2019MNRAS.487.1178P}.
Our non-rotating models are detectable to similar distances. 
For a
model with strong SASI activity, \citet{2016ApJ...829L..14K} found
higher amplitudes that could allow detection out to
$\mathord{\sim}60\, \mathrm{kpc}$ with aLIGO. 
Our SASI-dominated
model s18np does not reach such high amplitudes, and it
remains to be investigated further under which conditions
the SASI could give rise to significantly stronger 
gravitational-wave emission. 

Several studies have already examined the detection prospects for
gravitational-wave emission from rotational bounce and
non-axisymmetric instabilities in rapidly rotating progenitors
\citep{2014PhRvD..90d4001A,2016PhRvD..93d2002G,2018MNRAS.475L..91T,2019arXiv190909730S}. Unsurprisingly,
the maximum detection distance for our rapidly-rotating model m39 is
similar to these cases.  It is interesting, however, that
such a strong signal can be obtained for observers
in the polar direction (who would not see any bounce signal)
and without the help of non-axisymmetric instabilities.
This complicates the interpretation of distant gravitational-wave 
events. A high signal strength alone
may not be a telltale sign for the physics of the explosion,
though it would suggest that rapid rotation is involved. In particular,
the spectrogram of model m39, which undergoes
a strong, but otherwise ordinary convectively-aided
explosion,  is qualitative similar to the
case of the low-$T/|W|$ instability \citep{2019arXiv190909730S};
suggesting that there may be room for confusion between
these scenarios in the event of a gravitational-wave detection.

\section{Conclusions}
\label{sec:conclusion}
To survey the expected gravitational waveforms for supernova
explosions, modern simulations of 3D neutrino-driven explosions have
yet to explore a broader range of the CCSN progenitor parameter space.
In order to better cover the most promising regime for gravitational-wave 
detections, we particularly need more simulations of 
massive and rotating
progenitors, and longer simulations beyond the phase of peak
gravitational-wave emission. The systematics
of explosion and remnant properties also need to be explored
more thoroughly, in particular at the high-mass end of
the progenitor distribution.

In this paper, we therefore performed simulations of three different
high-mass progenitors, for a long enough
duration that we can produce robust predictions of the
gravitational-wave emission, the explosion energies, and the remnant
properties.  These progenitors include a rotating
$39\,M_\odot$ Wolf-Rayet star, a non-rotating $20\,M_\odot$ Wolf-Rayet star,
and an $18\,M_\odot$ red supergiant.

Both of the Wolf-Rayet models develop neutrino-driven explosions a
few hundred milliseconds after core bounce, whereas the
$18\,M_\odot$ model fails to explode without the help
of strong convective seed perturbations in agreement
with an earlier simulation of the same model \citep{2017MNRAS.472..491M}. 
These results further strengthen the case for
a regime of relatively early shock revival in
high-mass progenitors with massive cores and oxygen shells
\citep{2018ApJ...855L...3O,2019arXiv190904152B}.

The explosion energies for the two Wolf-Rayet stars
are still rising by the end of the simulation.
The energetics of the $20\,M_\odot$ model is on
track for a normal explosion with an energy
of $6 \times 10^{50}\, \mathrm{erg}$ and
at a post-bounce time of $1.2\, \mathrm{s}$, which
may yet increase to $\mathord{\sim} 10^{51}\, \mathrm{erg}$
after a few more seconds. By contrast, the rotating
$39\,M_\odot$ model has already reached
$10^{51}\, \mathrm{erg}$, which is
still increasing at a rate of $10^{51}\, \mathrm{erg}\,\mathrm{s}^{-1}$.
This demonstrates that the neutrino-driven
mechanism can produce powerful explosions and
not only account for underenergetic events.
Since the net accretion rate onto the
PNS is already small in this
model, the cycle of accretion and mass ejection
could presumably continue for several seconds
without black hole formation. Although
longer simulations are needed for verification,
we speculate that several $10^{51}\, \mathrm{erg}$
could be reached. This suggests that
neutrino heating might even play an important
subsidiary role in the early stages of
hypernova explosions by reviving the shock and delivering a
significant part of the hypernova explosion energy. 

Different from other rotating 3D explosion models
\citep{takiwaki_16,janka_16,summa},
in the $39\,M_\odot$ model the explosion occurs without
help from the spiral mode of the SASI or the low-$T/|W|$
instability, and is characterised by a bipolar outflow geometry
that is roughly aligned with the rotation axis.
As a result, the PNS
spin and kick remain aligned within $30^\circ$.
This suggests that rapid progenitor rotation 
should be explored further as a possible explanation
for spin-kick alignment, which cannot be accounted
for in non-rotating 3D models in a natural manner
\citep{wongwathanarat,2019MNRAS.484.3307M}.

We also examine the gravitational-wave emission of our models.  The
non-rotating models show the expected emission features, with a
spectrogram dominated by high-frequency f/g-modes, as well as a strong
low-frequency SASI mode in the case of the $20\,M_\odot$ progenitor. At
first glance, the $39\,M_\odot$ model comports with this familiar
picture as well, but reveals a new, interesting effect of rapid
rotation upon closer inspection. The f-mode frequency is more than a
factor of two lower than expected for the PNS
parameters. The reduced mode frequency can be explained by a negative
angular momentum gradient in the PNS surface region,
which counteracts the stabilising entropy gradient that normally sets
the f-mode frequency. The angular momentum gradient can be traced back
to differential rotation between the Si and O shell in the progenitor,
which is probably a generic phenomenon since angular momentum
transport across shell boundaries is generally less efficient
than within shells.
Attempts to constrain PNS
properties based on the mode frequencies
\citep{2019PhRvL.123e1102T} will need to address the potential
impact of rotation in future.

We estimate the distance to which the gravitational-wave emission of
our models could be detected by current ground-based
gravitational-wave detectors at design sensitivity, and by the future
Einstein Telescope.  Due to rapid rotation
and its high explosion energy, the
amplitude of the $39\,M_\odot$ model is large
enough for it to be detected to a
distance of almost 2\,Mpc in the Einstein Telescope.
Interestingly, the detection distance does not
depend very much on the observer direction, since
f/g-mode excitation contributes to the 
SNR to a similar degree as the
rotational bounce signal.

Our new models add further confidence that the neutrino-driven
scenario can explain a broader range of explosion properties and
produce rather powerful gravitational-wave signals in some cases.
Further simulations with complementary simulation methodologies, as
well as more extensive parameter studies are needed to corroborate
these optimistic findings.

\section*{Acknowledgements}

We thank D.~Aguilera-Dena, N.~Langer, and S.-C.~Yoon for
providing Wolf-Rayet star models.
JP is supported by the Australian Research Council (ARC) Centre of 
Excellence for Gravitational Wave Discovery (OzGrav), through project number 
CE170100004. BM is supported by ARC Future Fellowship FT160100035. Some of 
this work was performed on the Raijin supercomputer with the assistance of 
resources and services from the National Computational Infrastructure (NCI), 
which is supported by the Australian Government. Some of this work was 
performed on the OzSTAR national facility at Swinburne University of Technology. 
OzSTAR is funded by Swinburne University of Technology and the National 
Collaborative Research Infrastructure Strategy (NCRIS).


\bibliographystyle{mnras}
\interlinepenalty=10000
\bibliography{bibfile}


\appendix


\bsp	
\label{lastpage}
\end{document}